\documentclass[aps,pra,showpacs,superscriptaddress,amssymb,twocolumn]{revtex4-1}
\usepackage{hyperref}
\usepackage{graphicx}
\usepackage{appendix} 
\usepackage{color}\hypersetup{colorlinks=true,citecolor=blue,linkcolor=black}
\usepackage{bm,amsmath}
\usepackage{dcolumn}
\input{Qcircuit}

\newtheorem{thm}{Theorem}
\newtheorem{conjecture}[thm]{Conjecture}

\newcommand{\be}{\begin{equation}}
\newcommand{\ee}{\end{equation}}

\begin{document}

\rightline{\tt }

\vspace{0.2in}

\title{Sampling and scrambling on a chain of superconducting qubits}

\author{Michael R. Geller}
\affiliation{Center for Simulational Physics, University of Georgia, Athens, Georgia 30602, USA}
\date{\today}

\begin{abstract}
We study a circuit, the Josephson sampler,  that embeds a real vector into an entangled state of $n$ qubits, and optionally samples from it. We measure its fidelity and entanglement on the 16-qubit ibmqx5 chip. To assess its expressiveness, we also  measure its ability to generate  Haar random unitaries and quantum chaos, as measured by Porter-Thomas statistics and out-of-time-order correlation functions. 
The circuit requires nearest-neighbor CZ gates on a chain and is especially well suited for first-generation superconducting architectures.
\end{abstract}


\maketitle

\section{INTRODUCTION}

A common task in quantum computing is to embed large amounts of classical data into a quantum state of $n$ qubits. Large means more than $n$ bits, so the data must be input through a circuit with adjustable parameters. This has application to quantum machine learning \cite{BiamonteArxiv161109347}, where the data of interest (e.g., images) are kilobytes in size or larger, and also to quantum simulation, where the data can be used as variational parameters \cite{PeruzzoNatComm14,OMalleyPRX16,KandalaNat17}.  And inputting pseudorandom classical data into such a circuit can be used to approximate Haar random unitaries \cite{EmersonSci03,Brandao12080692,Brandao160500713}, which have wide application in quantum computing and quantum information \cite{EmersonSci03,DiVincenzoPRL04,HaydenCMP04,LumPRA16,DupuisProcRoySocA13}.

In this work we will study the performance of a practical embedding circuit---the {\em Josephson sampler}---on the IBM Quantum Experience ibmqx5 device, which has 16 transmon qubits. We study samplers up to size $n\!=\!6$. The circuit acts on a 1d chain of qubits with nearest-neighbor CNOT or CZ gates, and has a layered construction 
\begin{equation}
U = U_{L} \cdots U_2 U_1
\end{equation}
with $L$ layers, as shown below.

\begin{small}
\begin{center}
\mbox{
\Qcircuit @C=1em @R=.7em {
&&&&&&&&&&&&&&&&&&&& {} \\
&&&&&&&&&&&&&&&&&&&& {} \\
&&&\lstick{1 \ \ }  &  \gate{u} & \ctrl{1}  &  \qw & \qw & \qw                   &  \gate{u} & \ctrl{1}  &  \qw & \qw & \qw     \\
&&&\lstick{2 \ \ }  &  \gate{u} & \control  \qw  &   \gate{u} & \ctrl{1}       & \qw &  \gate{u} & \control  \qw  &   \gate{u} & \ctrl{1} & \qw  \\
&&&\lstick{3 \ \ }  &  \gate{u} & \ctrl{1}  &   \gate{u} & \control  \qw       & \qw &   \gate{u} & \ctrl{1}  &   \gate{u} & \control  \qw  & \qw   \\
&&&\lstick{4 \ \ }  &  \gate{u} & \control  \qw  &   \gate{u} & \ctrl{1}       & \qw &   \gate{u} & \control  \qw  &   \gate{u} & \ctrl{1} & \qw && \cdots   \\
&&&\lstick{5 \ \ }  &  \gate{u} & \ctrl{1}  &   \gate{u} & \control  \qw       & \qw &  \gate{u} & \ctrl{1}  &   \gate{u} & \control  \qw  & \qw  \\
&&&\lstick{6 \ \ }  &  \gate{u} & \control  \qw  &   \gate{u} & \ctrl{1}       & \qw &   \gate{u} & \control  \qw  &   \gate{u} & \ctrl{1} & \qw    \\
&&&&&&&&&&&&&&&&&&&& {} \\
&&&\lstick{\ \ \vdots \ \ }  && \vdots &&&&& \vdots &  \\
&&&&&&&&&&&&&&&&&&&& {} \\
&&&&&&&&&&&&&&&&&&&& {} \\
&&&&&&&&&&&&&&&&&&&& {} \\
&&&&&&&&&&&&&&&&&&&& {} \\
&&&&& U_1 &&&&& U_2 &&&&&&&&& {} \\
&&&&&&&&&&&&&&&&&&&& {} \\
&&&&&&&&&&&&&&&&&&&& {} 
\gategroup{3}{5}{12}{8}{1em}{--} 
\gategroup{3}{10}{12}{13}{1em}{--} 
\\
}}
\end{center}
\end{small}
Each gate $u$ is a single-qubit rotation gate
\begin{equation}
u(\theta,\phi) \equiv e^{-i \phi Z/2} \times e^{-i \theta Y/2},
\label{u definition}
\end{equation}
where the $y$ rotation is applied first and the $z$ rotation is done in software and carries no depth. The vertical two-qubit gates are CZ gates. Important features of the design are that the gateset is universal and the columns of CZs rapidly generate entanglement.  A sampler circuit with $L$ layers maps a real vector $x \in \mathbb{R}^{m}$ of dimension  
\begin{equation}
m = 2(2n-2)L
\end{equation}
to (ideally) a unitary $U \in {\rm U}(2^n).$ 
Additional details about the circuit are provided in Appendix A.

Our work is partly inspired by Aaronson and Arkhipov \cite{AaronsonTC13}, Boixo {\it et al.}~\cite{Boixo160800263},
Kandala {\it et al.}~\cite{KandalaNat17},
and Neill {\it et al.}~\cite{Neill170906678}.
Although our current objective is not quantum supremacy, which is unlikely on a chain, we will apply many of the techniques introduced in \cite{Boixo160800263} and \cite{Neill170906678}. 
A different but related problem of boson sampling with superconducting resonators was discussed in Refs.~\cite{PeropadrePRL16} and \cite{GoldsteinEtalPRB17}.
The Josephson sampler is an alternative to the hardware-efficient circuit introduced by Kandala {\it et al.}~\cite{KandalaNat17}, trading some performance for simpler portability. We will also extend previous work by measuring 4-point out-of-time-order correlation functions \cite{Maldacena150301409,SwinglePRA16}.
These probe the butterfly effect, a dynamical signature of quantum chaos, and information scrambling \cite{Hosur151104021,SwinglePRA16,Yao160701801}.

\section{FIDELITY}

\begin{figure}
\includegraphics[width=8.5cm]{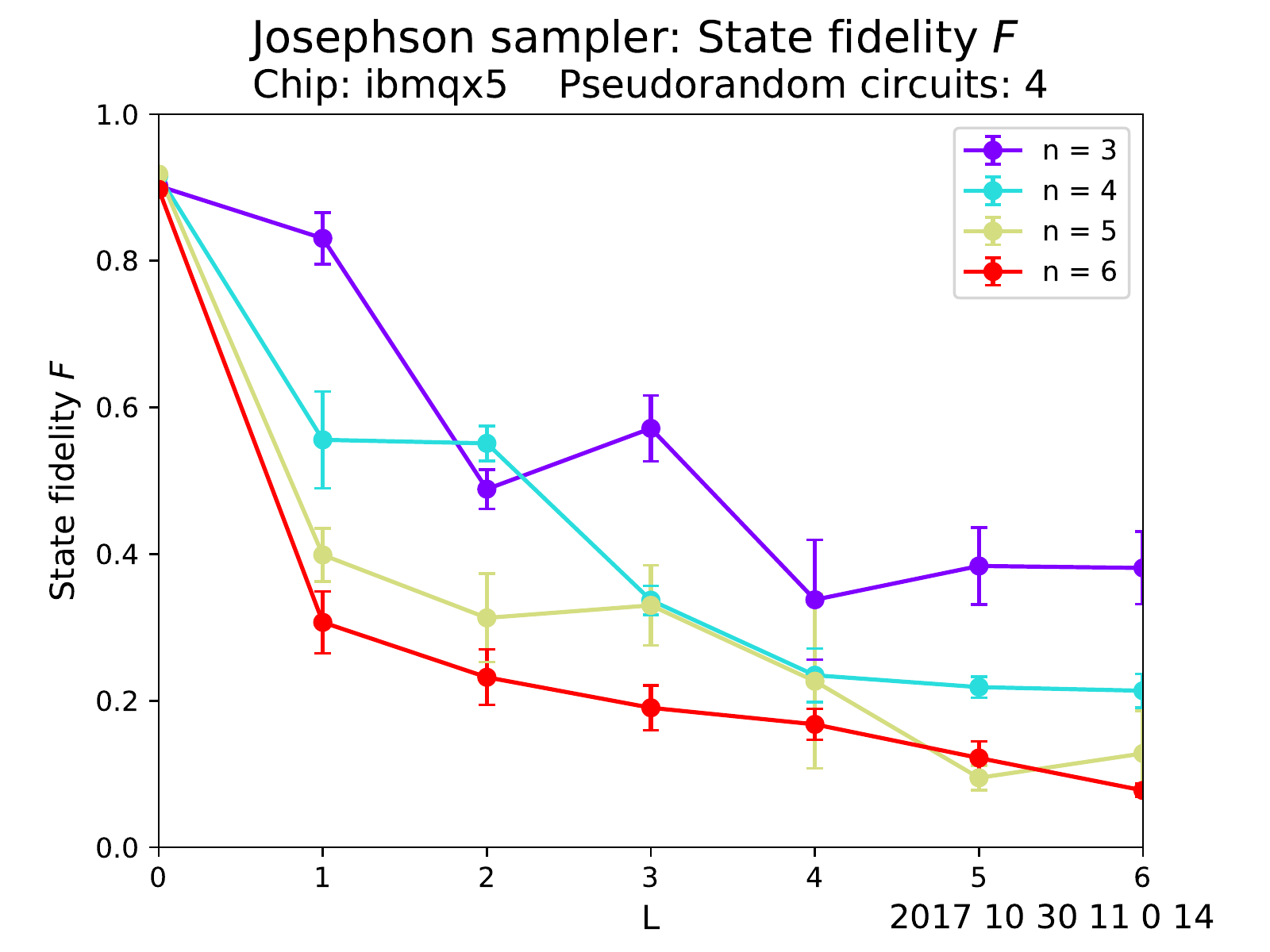} 
\caption{(Color online) Measured state fidelity (\ref{state fidelity definition}) averaged over $s$ pseudorandom circuits of size $(n,L),$ after application to initial state $|0\rangle^{\otimes n}.$ Each state fidelity measurement consists of $r$ Pauli expectation-value measurements. Here $s = 4$ and $r=8$. Error bars denote one standard error.}
\label{statefidelity}
\end{figure}

The Josephson sampler is designed for use on first-generation gate-based quantum computers, which are not error corrected. It is therefore critical to assess its performance on real devices. One way to measure the quality of a circuit implementation is to estimate the fidelity 
\begin{equation}
F  \equiv \bigg( \! {\rm Tr} \sqrt{\sqrt{\rho_{\rm t}}  \rho \sqrt{\rho_{\rm t}}} \bigg)^2 = {\rm Tr}  \rho \rho_{\rm t},
\label{state fidelity definition}
\end{equation}
of the resulting final state $\rho$ with its ideal pure target $\rho_{\rm t}.$ 
Given a classically precomputed $\rho_{\rm t}$, there is a randomized protocol to estimate $F$, by Flammia and Liu \cite{FlammiaPRL11}, which we find to converge very quickly. In Fig.~\ref{statefidelity} we plot the state fidelity versus number of sampler layers $L$. For each sampler size $(n,L)$, the average state fidelity over $s$ pseudorandom circuits is measured, with $s=4$. The Flammia-Liu protocol \cite{FlammiaPRL11} is based on an expansion of the density matrix in the Pauli basis, but requires expectation-value measurements for only a small number $r$ of the more probable ones. In Fig.~\ref{statefidelity} we used $r = 8$ Pauli operator samples.

Boixo {\it et al.}~\cite{Boixo160800263} introduced alternative information-theoretic tools to quantify circuit fidelity. Their approach starts with a classically precomputed ideal probability distribution $p_{\rm ideal}(x)$, with $x \in \{0,1\}^{\otimes n}$ a classical state, or an algorithm for computing it. The goal is to experimentally distinguish between the actual distribution $p_{\rm meas}(x)$ and the ideal one $p_{\rm ideal}(x)$. But we want to do this {\it without} an accurate estimate of $p_{\rm meas}(x),$ which would require $O(2^n)$ measurement samples.

The idea behind the method of Boixo {\it et al.}~\cite{Boixo160800263} is to focus on the rare, high-information events and their statistics. Given $p_{\rm ideal}(x)$ we define the {\it information} $s(x) \equiv \log_2 [1/p_{\rm ideal}(x)]$ contained in each classical state $x$, measured in bits. First we ask some purely theoretical questions about the ideal distribution $p_{\rm ideal}(x)$. For example, we can calculate the average information 
\begin{equation}
S_{\rm ideal}  \equiv \big\langle s(x) \big\rangle_{\rm ideal} = \sum_x  p_{\rm ideal}(x) \, s(x)
\end{equation}
when sampling from $p_{\rm ideal}(x)$, which is the classical Shannon entropy. We can also calculate the uniform average
\begin{equation}
S_{\rm unif} \equiv \big\langle s(x) \big\rangle_{\rm unif} = \frac{1}{N} \sum_x s(x), \ \ \ N = 2^n,
\end{equation}
which is larger than the entropy because it over-represents the rare, ${\rm prob} \! < \! 1/N$ high-information events.
Therefore the difference
\begin{equation}
S_{\rm unif} - S_{\rm ideal}
\end{equation}
is sensitive to the statistics of rare events. 

\begin{figure}
\includegraphics[width=8.5cm]{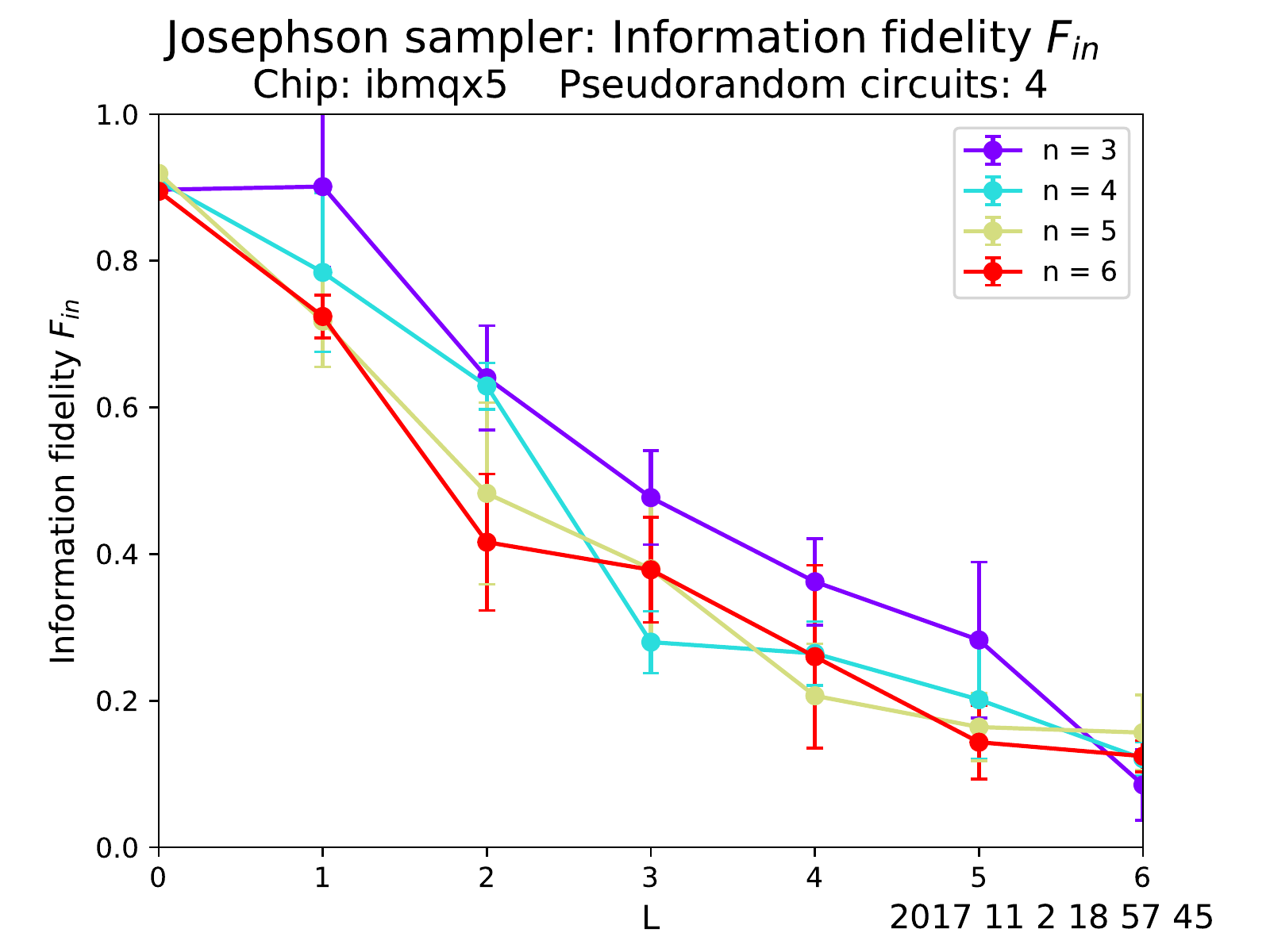} 
\caption{(Color online) Measured information fidelity (\ref{information fidelity definition}) averaged over $s$ pseudorandom circuits (the same circuits used in Fig.~\ref{statefidelity}).}
\label{informationfidelity}
\end{figure}

In the information-theoretic approach the quantity actually measured is the cross-entropy
\begin{equation}
H_{\rm c}  \equiv \big\langle s(x) \big\rangle_{\rm meas} = \sum_x  p_{\rm meas}(x) \, s(x).
\end{equation}
Note that it is not necessary to explicitly reconstruct $p_{\rm meas}(x)$, only to sample from it. So a possible fidelity measure is
\begin{equation}
F_{\rm in} \equiv 
\frac{S_{\rm unif} - H_{\rm c}}{S_{\rm unif} - S_{\rm ideal}},
\label{information fidelity definition}
\end{equation}
which was recently used by Neill {\it et al.}~\cite{Neill170906678}. Our measured values of $F_{\rm in}$ are shown in Fig.~\ref{informationfidelity}. An advantage of information fidelity measurement is that it only requires a single $H_{\rm c}$ estimate per sampler circuit, an $r$-fold reduction relative to state fidelity estimation.  (And going beyond the small circuits studied here, cross-entropy estimation should scale better than fidelity estimation \cite{Boixo160800263}.) The information fidelity data is also less noisy than the state data. However a possible weakness of definition (\ref{information fidelity definition}) is that there is not much $n$ dependence in Fig.~\ref{informationfidelity}, which does not seem physical.

\begin{figure}
\includegraphics[width=8.5cm]{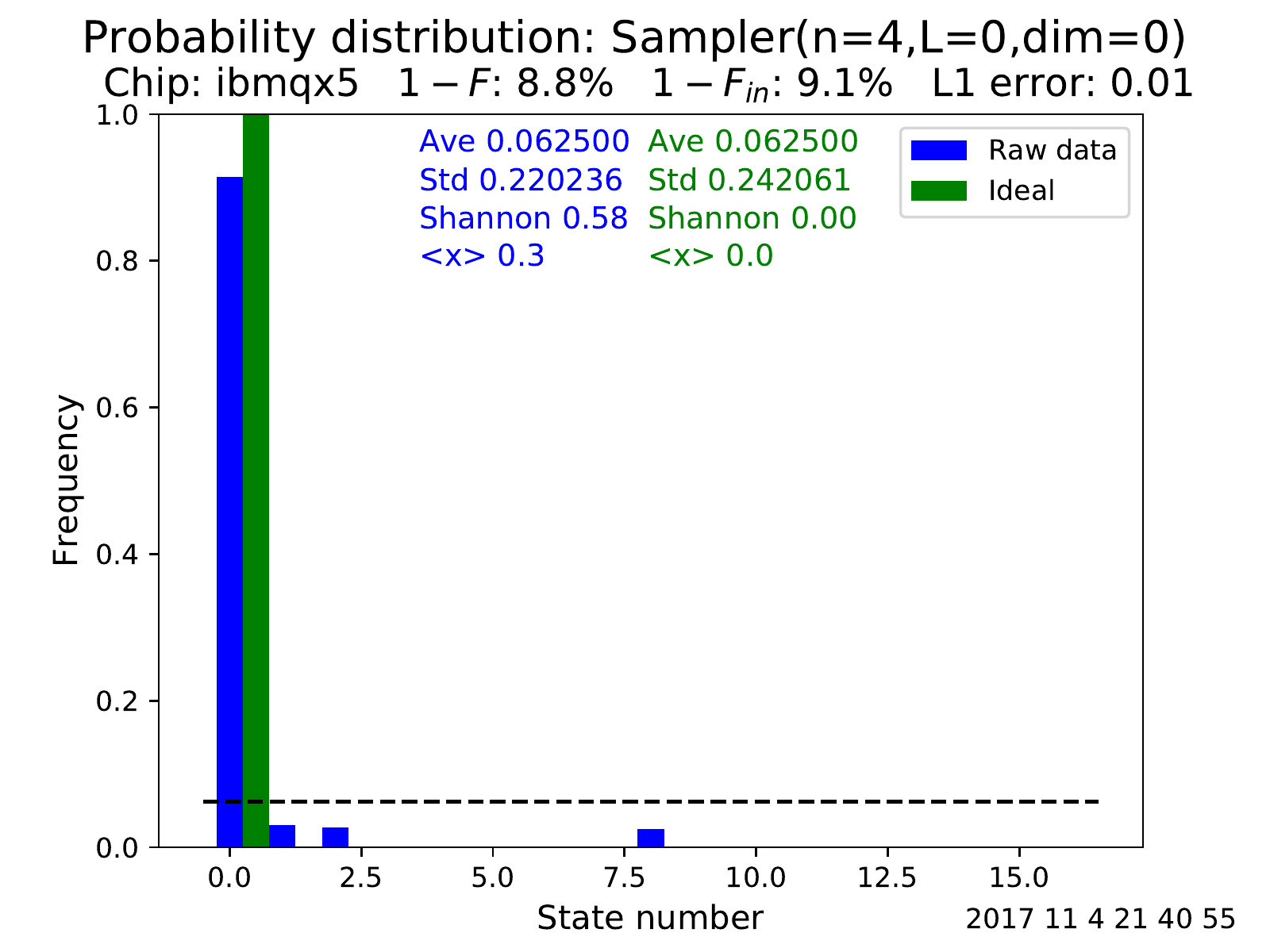} 
\caption{(Color online) Measured (blue) and simulated ideal (green) probability distributions resulting from state preparation immediately followed by readout in the computational basis.}
\label{n4L0}
\end{figure} 

\begin{figure}
\includegraphics[width=8.5cm]{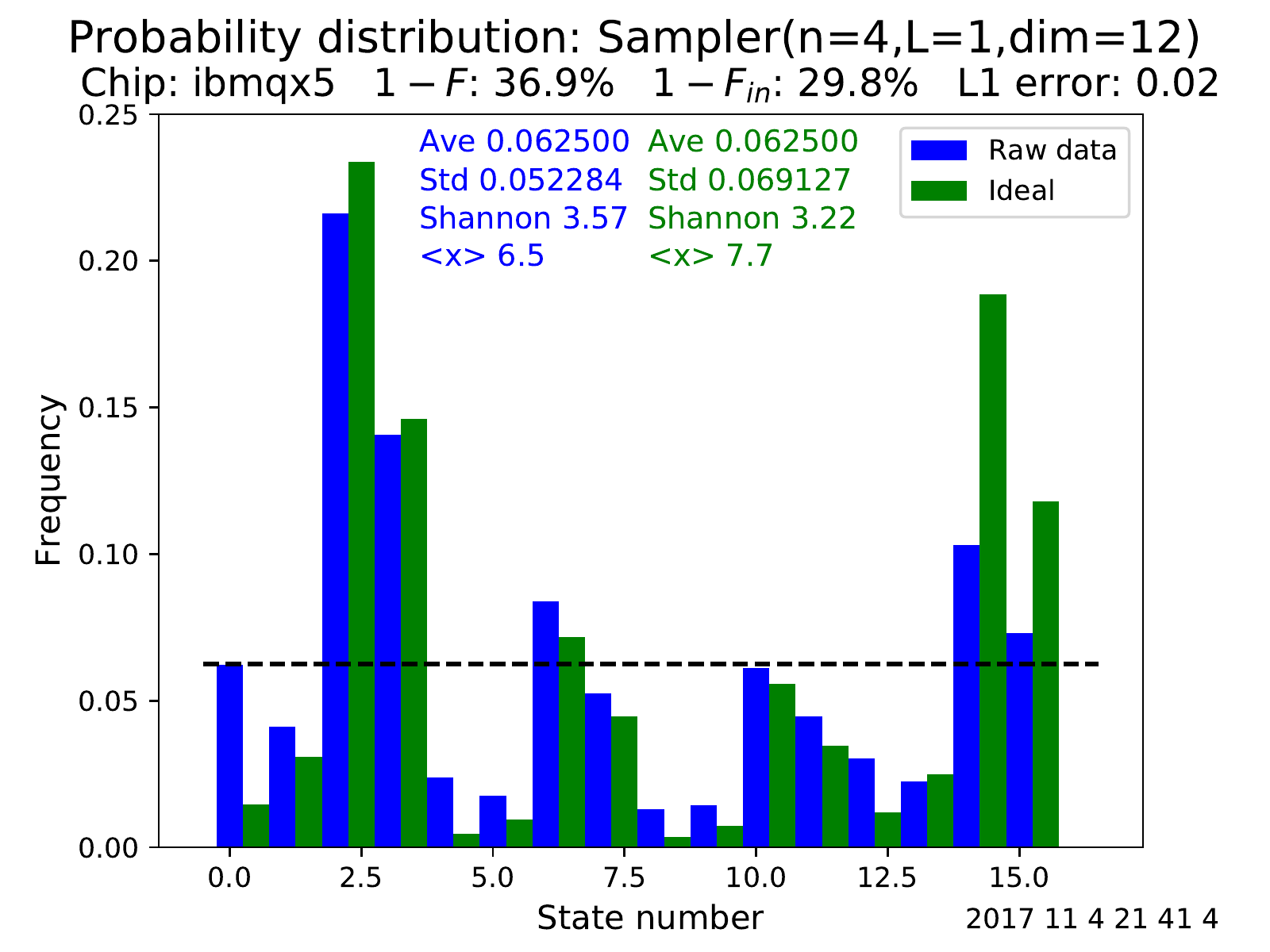} 
\caption{(Color online) Measured (blue) and ideal (green) probability distributions for a pseudorandom sampler of size $(n\!=\!4, L\!=\!1)$, which embeds a vector of dimension $m=12$. The dashed line is $1/N$.}
\label{n4L1}
\end{figure} 

\begin{figure}
\includegraphics[width=8.5cm]{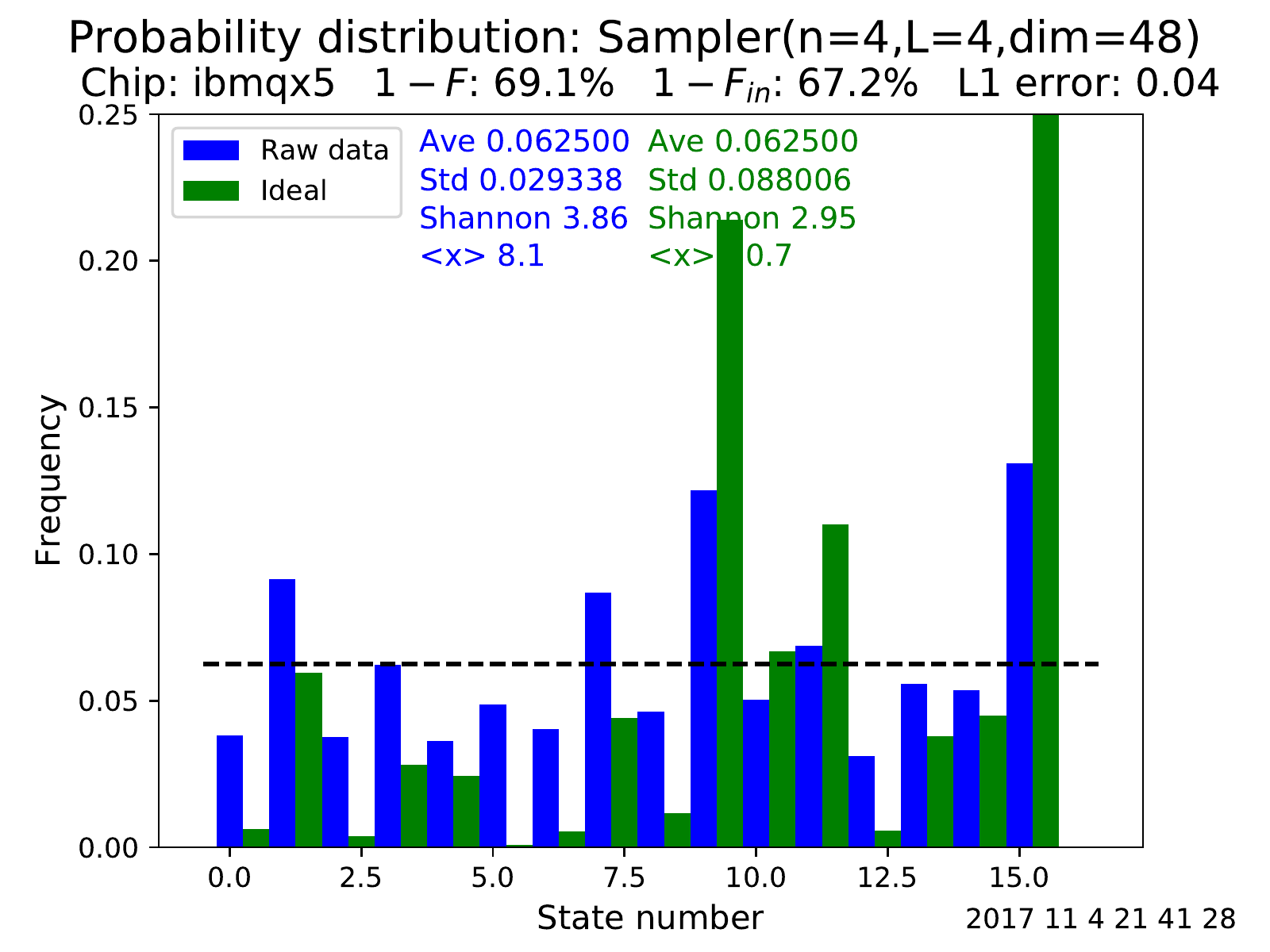} 
\caption{(Color online) Measured (blue) and simulated (green) probability distributions after four layers.}
\label{n4L4}
\end{figure}

\begin{figure}
\includegraphics[width=8.5cm]{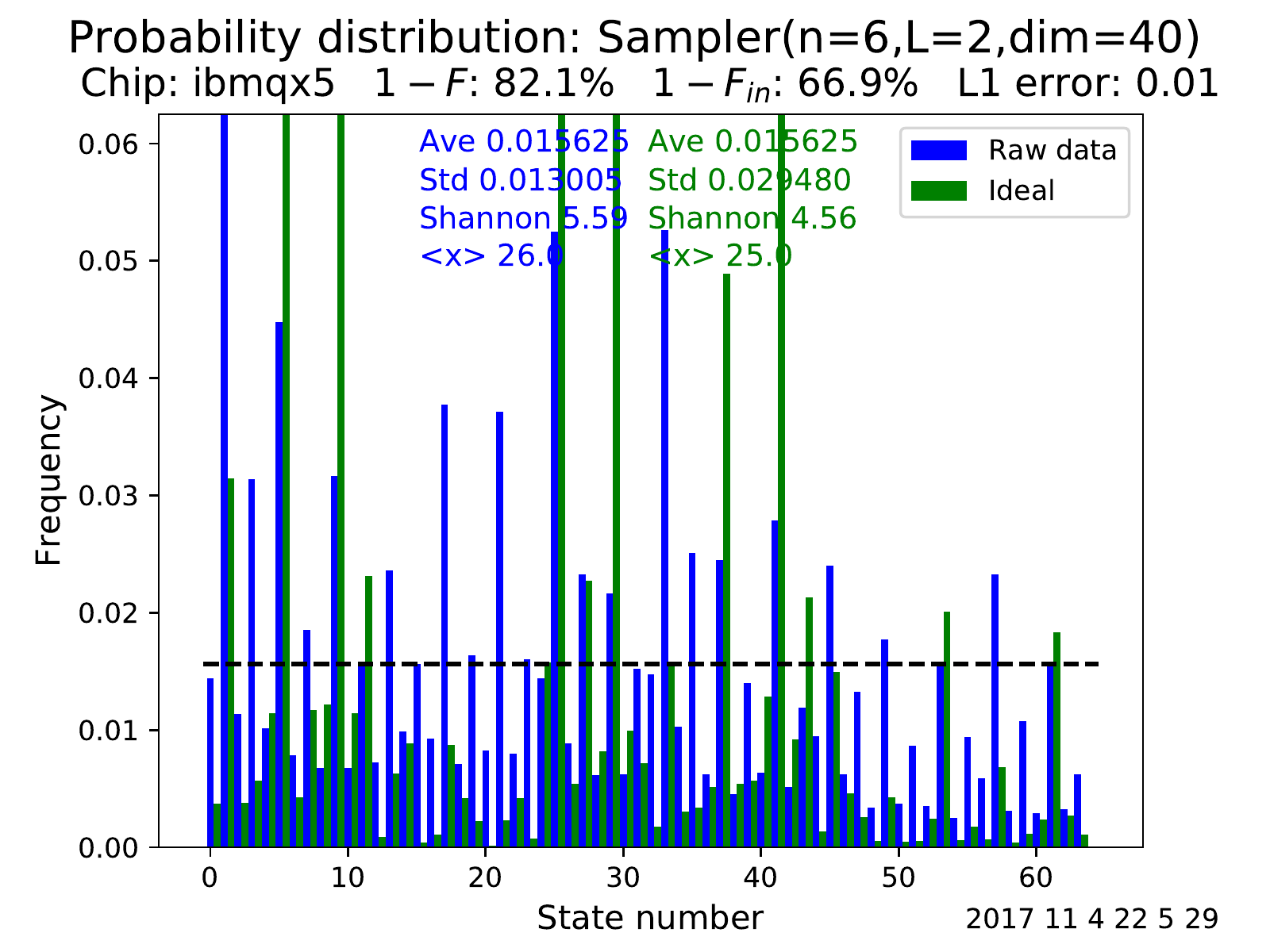} 
\caption{(Color online) Measured (blue) and ideal (green) probability distributions for a 6-qubit sampler after two layers.}
\label{n6L2}
\end{figure}

The remarkable similarity between Figs.~\ref{statefidelity} and \ref{informationfidelity} was predicted in \cite{Boixo160800263}. Here we explain the connection in a different but equivalent way: Let's assume a depolarizing error model for the physical density matrix,
\begin{equation}
\rho = (1-\epsilon) \rho_{\rm ideal} + \epsilon \rho_{\rm unif}, \ \ 
\rho_{\rm unif} = I / N,
\ \
0 \le \epsilon \le 1.
\end{equation}
In this model
$p_{\rm meas}(x) = (1-\epsilon) \, p_{\rm ideal}(x) + \epsilon  \, p_{\rm unif} $
and, by a direct calculation, $F_{\rm in} = 1 - \epsilon. $ We can say that the information fidelity is a direct measurement of the depolarizing error $\epsilon$. Evaluating (\ref{state fidelity definition}) for the same error model gives 
\begin{equation}
1 - F =  (1 - F_{\rm in} ) \bigg(1 - \frac{1}{N} \bigg).
\label{relation}
\end{equation}
Apart from a $1/N$ correction, where $N \! = \! 2^n$, the state and information fidelities are identical for a depolarizing channel. The relation (\ref{relation}) is not expected to hold for other error models, but in the chaotic regime the physical errors become symmetrized, due to the action of a Haar random (or even 2-design) circuit, to a depolarized form \cite{EmersonSci07,MagesanPRA12,Boixo160800263}. This is how we interpret the overall agreement between  Figs.~\ref{statefidelity} and \ref{informationfidelity}. 

\section{SAMPLING}

Figure~\ref{n4L0} shows the probability distribution $p(x)$ after preparing the state $|0000\rangle$ and then immediately measuring. This serves to measure the readout errors and explain the subsequent figures. Here $x \in \{ 0, 1, \cdots, N-1\}$ labels the $N$ classical states, with $N = 2^n$. The dashed line is $1/N$.  Each probability distribution $p(x)$ is separately characterized in four ways: The average probability or event frequency ({Ave}), which is always $1/N$, the width of the frequency distribution as measured by the standard deviation ({Std}), the classical entropy ({Shannon}), and the mean index ($\langle x \rangle$).  And we quantify the difference between $p_{\rm meas}$ and $p_{\rm ideal}$ in three different ways: The state fidelity loss $1-F$, the cross-entropy error $1-F_{\rm in}$, and the L1 error, 
\begin{equation}
\frac{1}{N} \sum_x \big| p_{\rm meas} - p_{\rm ideal} \big|,
\end{equation}
which is the L1 distance divided by $N$. After one layer, Fig.~\ref{n4L1}, the probability distributions begin to spread out, and after four layers, Fig.~\ref{n4L4}, they appear to be highly scrambled but poorly correlated with each other. A probability distribution from the 6-qubit sampler is shown in Fig.~\ref{n6L2}.

\section{ENTROPY AND ENTANGLEMENT}

\begin{figure}
\includegraphics[width=8.5cm]{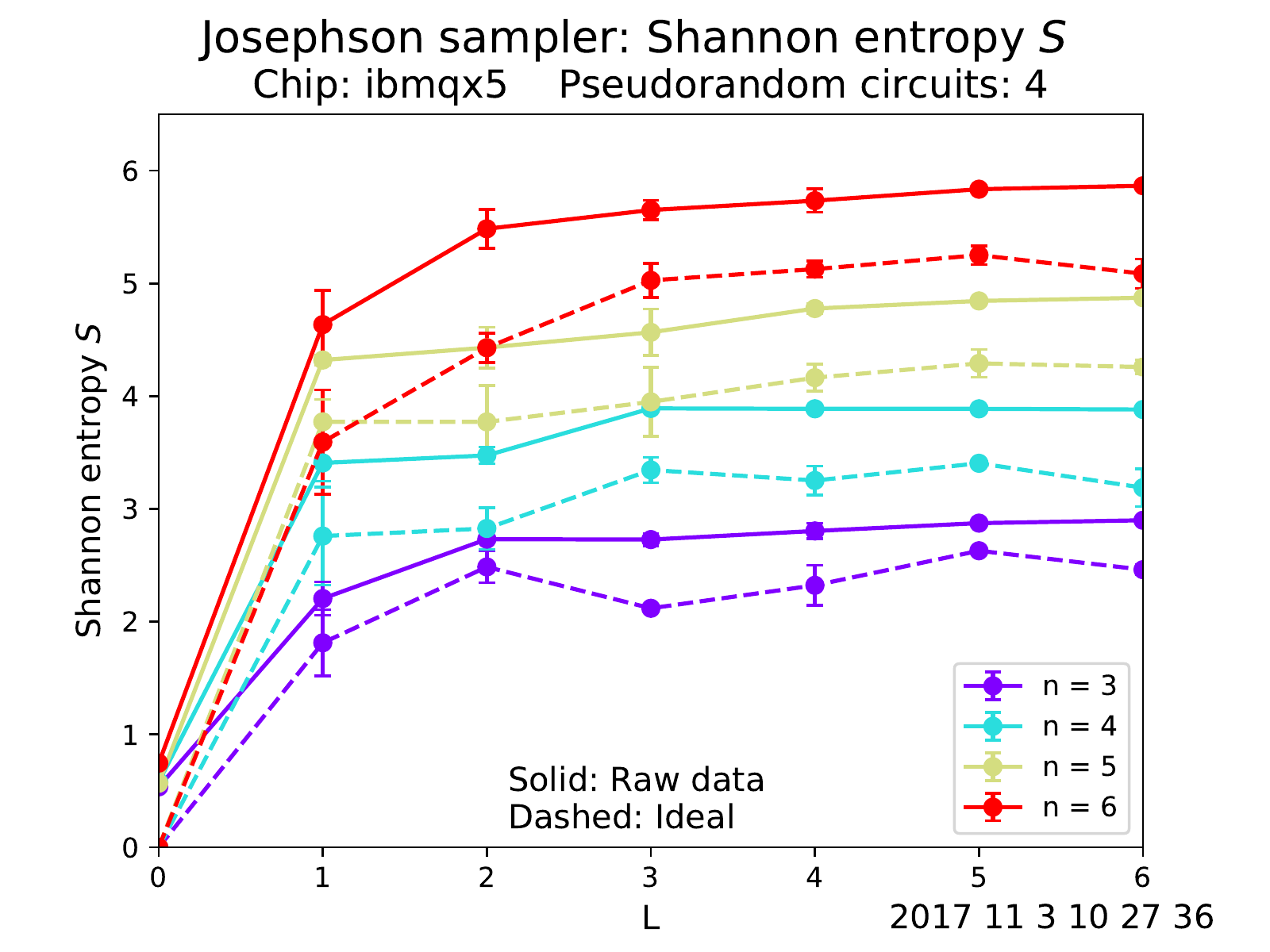} 
\caption{(Color online) Measured (solid curves) and simulated (dashed) entropy versus number of sampler layers. The classical simulations includes the ideal circuit unitaries but no gate errors or decoherence.}
\label{entropy}
\end{figure}

It's also interesting to measure the classical entropy generated by the sampler as a function of $L$; this is shown in Fig.~\ref{entropy}. Note that measured entropies almost reach their maximum of $n$ bits, but that the ideal entropies are about one bit short (we will discuss this in Sec.~\ref{HAAR TYPICALITY}). The data suggest that $n-1$ of the $n$ bits of generated entropy came from the unitary dynamics of the circuit, and that decoherence does not have a strong effect on the entropy production (relative to its effect on fidelity). Perhaps this is because the entropy is already increasing very rapidly due to the unitary evolution.

\begin{figure}
\includegraphics[width=8.5cm]{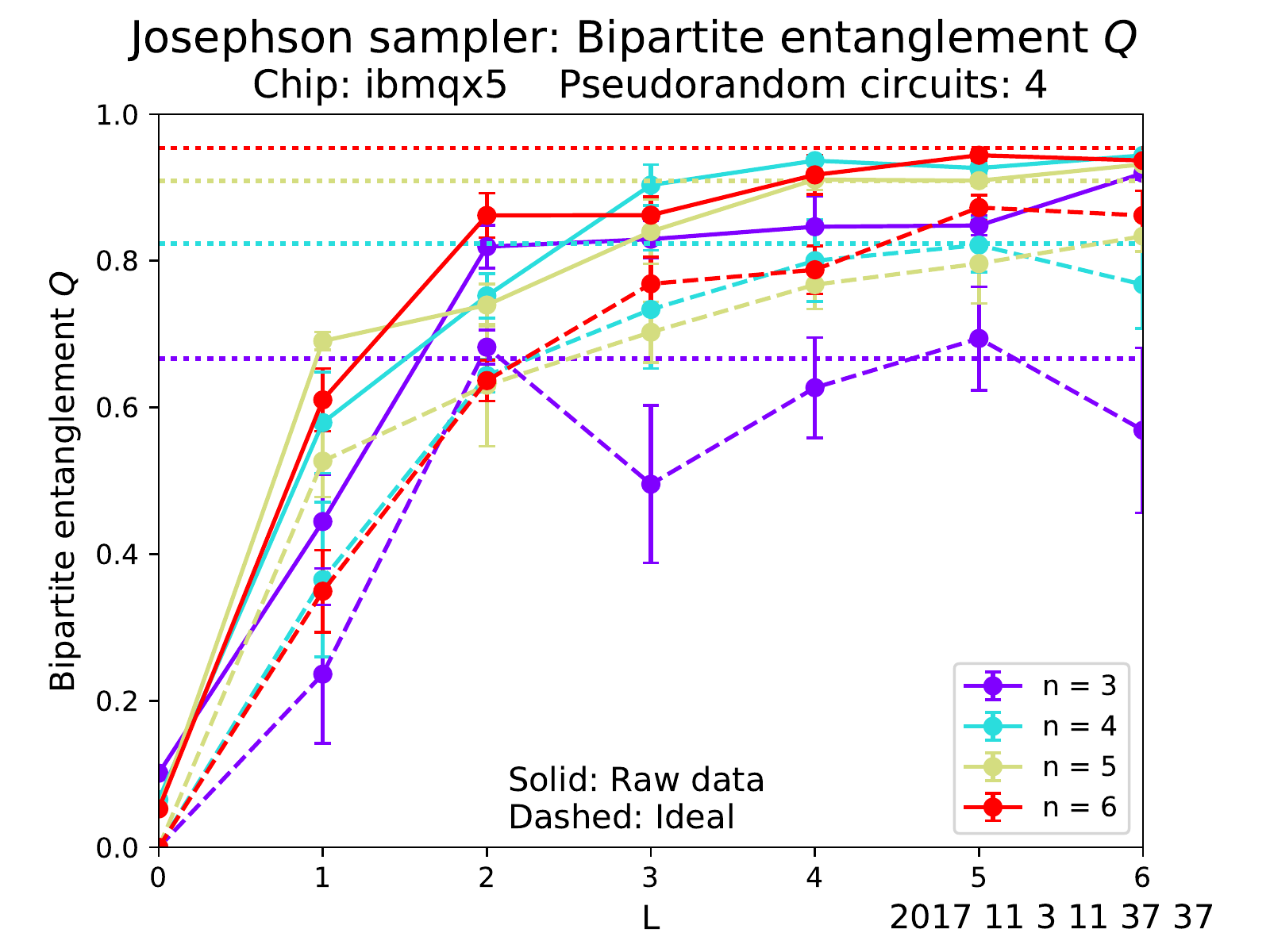} 
\caption{(Color online) Measured entanglement (\ref{bipartiteEntanglementDefinition}) averaged over $s$ pseudorandom circuits (solid curves), along with their ideal classically simulated values (dashed). The horizontal dotted lines show the Haar averages (\ref{Qhaar}).}
\label{bipartiteEntanglement}
\end{figure}

There are several entanglement measures we will study, most of which are forms of subsystem quantum entropies, where the subsystem is one of the qubits on the chain. It is simple to measure these quantities here because we can directly reconstruct the single-qubit reduced density matrix
\begin{equation}
\rho_i = {\rm Tr}_{j\neq i} (\rho)
\label{reduceddensitymatrix}
\end{equation}
by tomography, which we carry out, one qubit at a time, tracing over (or not reading out) the other qubits. 
The average bipartite entanglement \cite{EmersonSci03}
\begin{equation}
Q = 2(1- \overline{\gamma}),
\ \ \ 
\overline{\gamma} = \frac{1}{n} \sum_{i=1}^n \gamma_i,
\ \ \
\gamma_i = {\rm Tr} (\rho_i^2),
\label{bipartiteEntanglementDefinition}
\end{equation}
is plotted in Fig.~\ref{bipartiteEntanglement}. Here $\gamma_i$ is the purity of the reduced density matrix (\ref{reduceddensitymatrix}). For each sampler size $(n,L),$ $Q$ is measured for a set of $s$ pseudorandom circuits, each requiring $3n$ measurements. The factor of 3 comes from the tomography operations required to measure each $\gamma_i$, and there are $n$ of them to measure. (The data in Fig.~\ref{bipartiteEntanglement} required the implementation and measurement of 1512 distinct circuits.) The results are encouraging, given that a high degree of entanglement is obtained after two layers. We can also look for an $n$-dependence to how quickly entanglement is achieved: We would normally expect that longer chains would entangle more slowly (after more circuit depth) than smaller ones, but the data shows strikingly little $n$ dependence.

A weakness of the entanglement measure (\ref{bipartiteEntanglementDefinition}) is that it is based on purity loss, which is also caused by decoherence. Thus, some of the entanglement we are detecting is really decoherence. One indication of this is that in the $n\!=\!3$ and $n\!=\!4$ cases, the measured values exceed their Haar averages  \cite{EmersonSci03}
\begin{equation}
\langle Q \rangle_{\rm Haar} = \frac{N-2}{N+1}, \ \ \ N = 2^n,
\label{Qhaar}
\end{equation}
which are listed in Table \ref{QhaarTable} and plotted as horizontal dotted lines in Fig.~\ref{bipartiteEntanglement}.
While exceeding the Haar average is theoretically possible, it is also exponentially unlikely due to measure concentration \cite{Hayden0407049}.  A second indication comes from comparing the data (solid curves) to a simulation (dashed curves) of the idealized problem with no gate errors and no decoherence. The measured data are typically 5-10\% higher in $Q$. So we might conclude that as much as 10\% of the $Q$ we are observing is not genuine entanglement. This would still be quite impressive given how much entanglement {\em is} generated. However it is easy to get misled by this entanglement measure because it can remain finite after the state has decohered (although it should eventually vanish when qubits relax to their nonentangled ground state $|0\rangle^{\otimes n}$).

\begin{table}[htb]
\centering
\caption{Haar average values of $Q$ from (\ref{Qhaar}).}
\begin{tabular}{|c|c|c|c|c|}
\hline
 $n$ & 3 &  4 & 5 & 6 \\
\hline
$\langle Q \rangle_{\rm Haar}$ & 0.667 & 0.824 &  0.909 & 0.954 \\
\hline 
\end{tabular}
\label{QhaarTable}
\end{table}

\begin{figure}
\includegraphics[width=8.5cm]{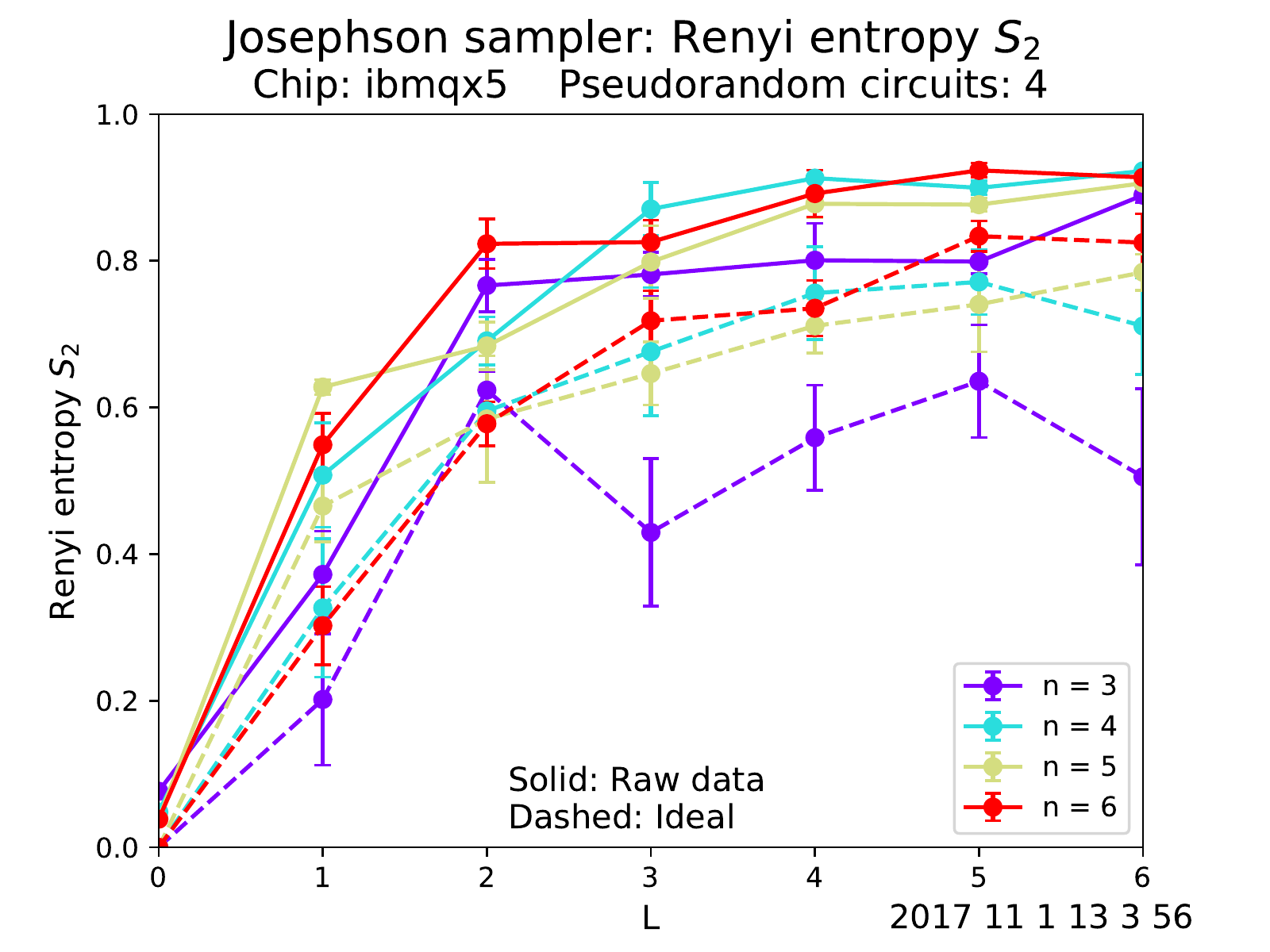} 
\caption{(Color online) Measured (solid curves) and simulated (dashed) R\' enyi entropy $S_2$, using the same data presented in Fig.~\ref{bipartiteEntanglement}.}
\label{renyi}
\end{figure} 

Another entanglement measure derived from the single-qubit purity $\gamma_i$ is the average second R\' enyi  entropy
\begin{equation}
S_2  = - \frac{1}{n} \sum_{i=1}^n \log_2 \gamma_i,
\label{renyi2 definition}
\end{equation}
which is plotted in Fig.~\ref{renyi}. The second R\' enyi entropy for a single qubit  satisfies $0 \le S_2 \le 1$, the same bounds as $Q$. Apart from the low-entanglement (small $L$) regime, we find that $S_2$ is almost identical to $Q$. The reason for the similarity is that the fluctuations in the $\gamma_i$ are small, at least for $n>3$ (note the small error bars in Fig.~\ref{bipartiteEntanglement}), so we can approximate $\gamma_i \approx \overline{\gamma}$. Then linearizing $S_2$ about a reference purity $\gamma_0$ we have
\begin{equation}
S_2  \approx  - \log_2(\gamma_0) + \frac{ \gamma_0 - \overline{\gamma}}{ \gamma_0 \ln 2} 
= a - b \,  \overline{\gamma},
\label{linearized entropy}
\end{equation}
where
$a = \log_2(1/ \gamma_0) + 1/\ln(2)$ 
and
$b = 1 / (\gamma_0 \ln 2)$.
The linearized R\' enyi entropy (\ref{linearized entropy}) would be identical to $Q$ if $a \! = \!  b  \! = \! 2$. Although this condition is not met for any choice of $\gamma_0$, the choice $\gamma_0 \! = \! 0.7$ leads to $a \! = \! 1.96$ and $b \! = \! 2.06$, which is quite close. Therefore we can view 
the average bipartite entanglement (\ref{bipartiteEntanglementDefinition}) as a linearized second R\' enyi entropy, corrected to assure the exact behavior in the limits $\overline{\gamma} \rightarrow 1/2$ and $\overline{\gamma} \rightarrow 1$.

\begin{figure}
\includegraphics[width=8.5cm]{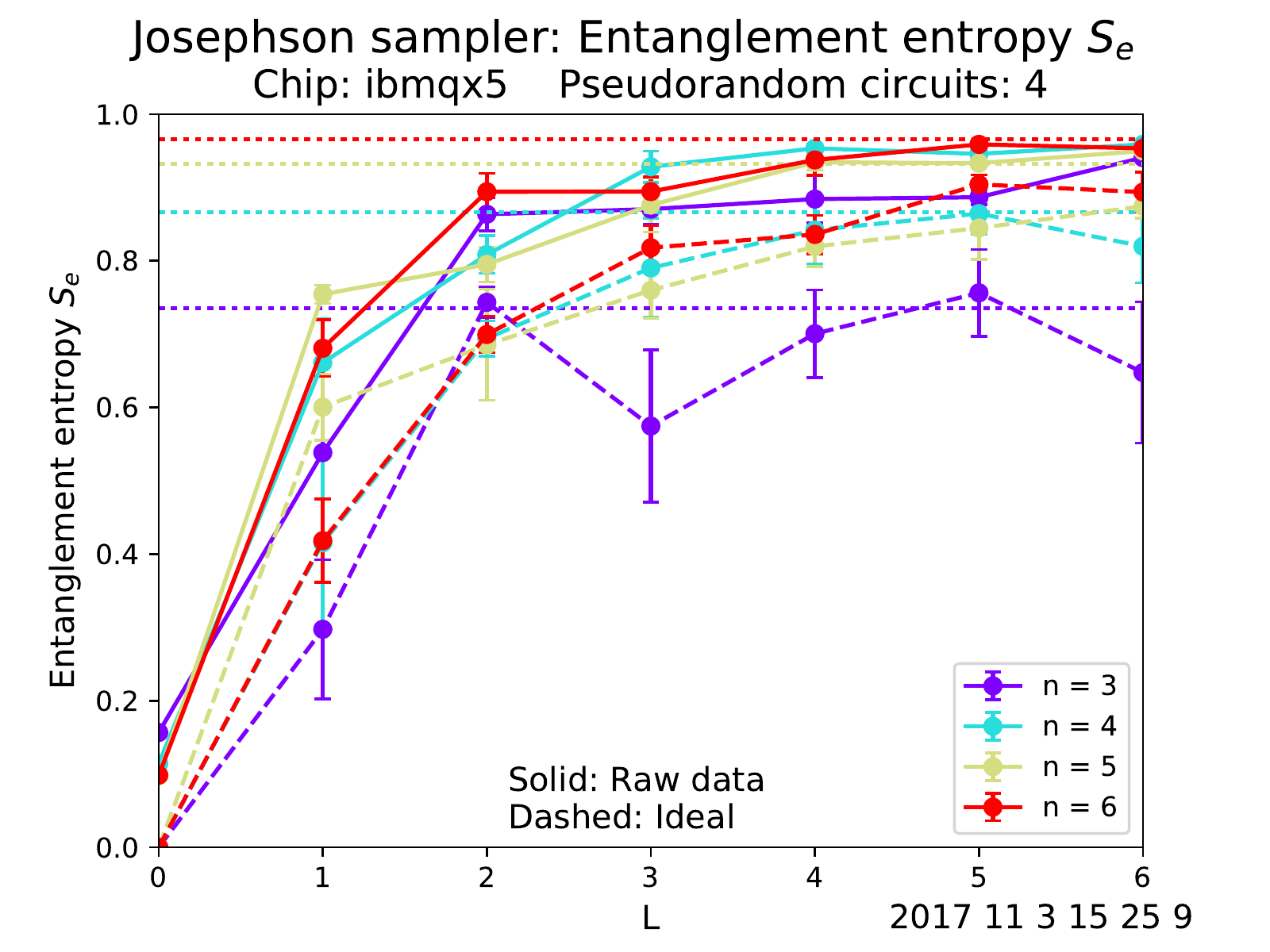} 
\caption{(Color online) Measured (solid curves) and simulated (dashed) entanglement entropy, using the same data as in Fig.~\ref{bipartiteEntanglement}.}
\label{entanglementEntropy}
\end{figure} 

Finally we measure an average entanglement entropy 
\begin{equation}
S_{\rm e}  = - \frac{1}{n} \sum_{i=1}^n {\rm Tr}( \rho_i \log_2 \rho_i),
\label{entanglementEntropyDefinition}
\end{equation}
measured in bits. This is shown in Fig.~\ref{entanglementEntropy}. As a point of reference we also plot the Haar-averaged entanglement entropy \cite{PagePRL93}
\begin{equation}
\langle S_{\rm e} \rangle_{\rm Haar} = \frac{1}{\ln 2} \bigg[ \sum_{k=d_{B}+1}^{ d_{A} d_{B} } \frac{1}{k} \ - \  \frac{d_{A}-1}{2 d_{B}} \bigg]
\label{averageEntanglementEntropy}
\end{equation}
for a bipartition of the chain into subsystems $A$ and $B$ with Hilbert space dimensions $d_{A} \! = \! 2$ and $d_{B} \! = \! 2^{n-1}.$ We note that the ideal entropies do not quite reach their Haar average values, while the measured ones exceed them. The entanglement entropy (in different contexts) was also measured in Refs.~\cite{NeillNatPhys16} and \cite{Xu170907734}.

\section{HAAR TYPICALITY}
\label{HAAR TYPICALITY}

Next we use the Josephson sampler to approximate Haar random unitaries. We will do this in a standard way, by inputting pseudorandom vectors $x \in \mathbb{R}^m$ of rotation angles into the circuit, thereby making the circuit itself pseudorandom. Our goal is to experimentally measure the quality of the resulting random unitaries and the scrambling of quantum information they produce. There are different aspects of the random unitaries that we might want to assess.  A measure of quality for an {\em ensemble} of random unitaries can be defined by regarding them as $\epsilon$-approximate $k$-designs \cite{Brandao160500713} and determining $\epsilon$  as a function of $k$. However this might be more useful as a theoretical tool, where one can try to calculate $\epsilon$ versus $k$. Instead we will focus on a different aspect of the random unitaries: whether they are Haar {\em typical}. 

\begin{figure}
\includegraphics[width=8.5cm]{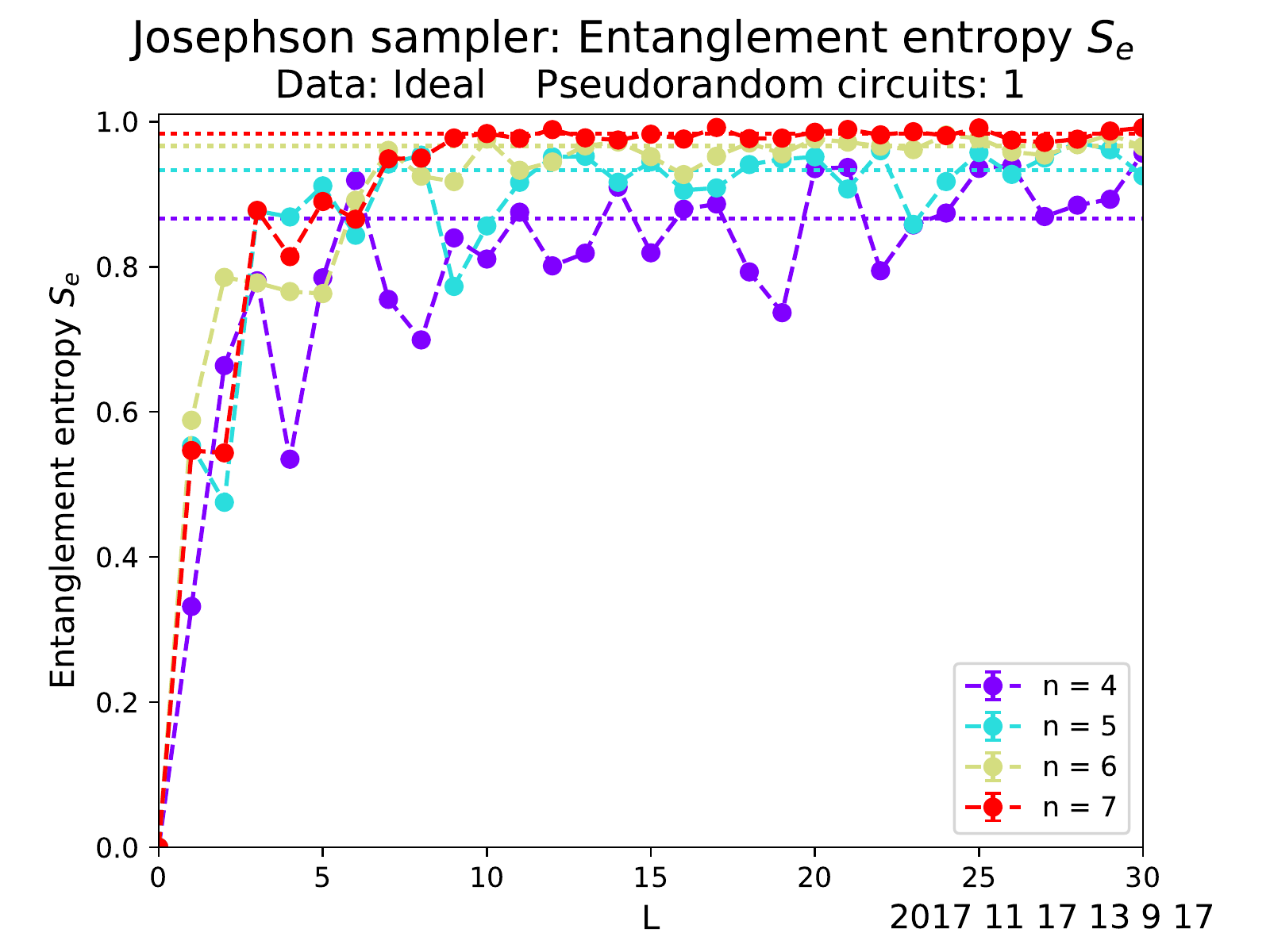} 
\caption{(Color online) Ideal entanglement entropy (\ref{entanglementEntropyDefinition}) for a single pseudorandom circuit (dashed curves). The horizontal dotted lines show the Haar averages (\ref{averageEntanglementEntropy}). This is the same as Fig.~\ref{entanglementEntropy}, except that here we use only one circuit, explore larger $n$ and $L$, and only include simulation results.}
\label{entanglementEntropyLarge}
\end{figure} 

In this section we will investigate an aspect of random unitaries inspired by the following proposition:

\begin{conjecture}
[Haar typicality]
Let $C$ be a quantum circuit on $n$ qubits for generating approximate Haar-random unitaries $U \in {\rm U}(2^n).$ Then it is possible to experimentally validate $C$ with only a single random instance. 
\label{HaarTypicallityConjecture}
\end{conjecture}

In practical terms we are saying that we can experimentally test whether a given quantum circuit is random or not. The complexity of $C$ does not matter. The intuition for a single-instance diagnosis comes from the exponentially sharp distributions of Lipschitz-continuous functions of $U \in {\rm U}(2^n)$ about their Haar averages \cite{Hayden0407049}. This means that with overwhelming probability, every $U$, when drawn uniformly from the group, will be chaotic and possess (some) universal attributes equal to their Haar-averaged values. The attributes are universal in the sense that they only depend on $n$. In the upside-down world of random unitaries, all $U$'s look alike, so in the error-free limit the Haar typicality conjecture would follow from the results of Hayden and coworkers \cite{Hayden0407049} on the concentration of entanglement entropy. Our main assertion, then, is that a practical typicality test is possible in the presence of small but finite gate errors and decoherence.  

Suppose we were to use the entanglement entropy (\ref{entanglementEntropyDefinition}) to diagnose Haar typicality. In Fig.~\ref{entanglementEntropyLarge} we plot the ideal $S_{\rm e}$ for a single pseudorandom sampler.  The entanglement entropy rapidly reaches the Haar-typical values, plotted as horizontal dotted lines. But, as we observe from Fig.~\ref{entanglementEntropy}, this measure is not sufficiently robust against gate errors and decoherence to diagnose Haar typicality. We can say that a necessary condition for a circuit to be Haar typical is a measured $S_{\rm e}$ at least as large as $\langle S_{\rm e} \rangle_{\rm Haar}$, but this condition is not sufficient. The bipartite entanglement (\ref{bipartiteEntanglementDefinition}) suffers from the same sensitivity to errors.

\begin{figure}
\includegraphics[width=8.5cm]{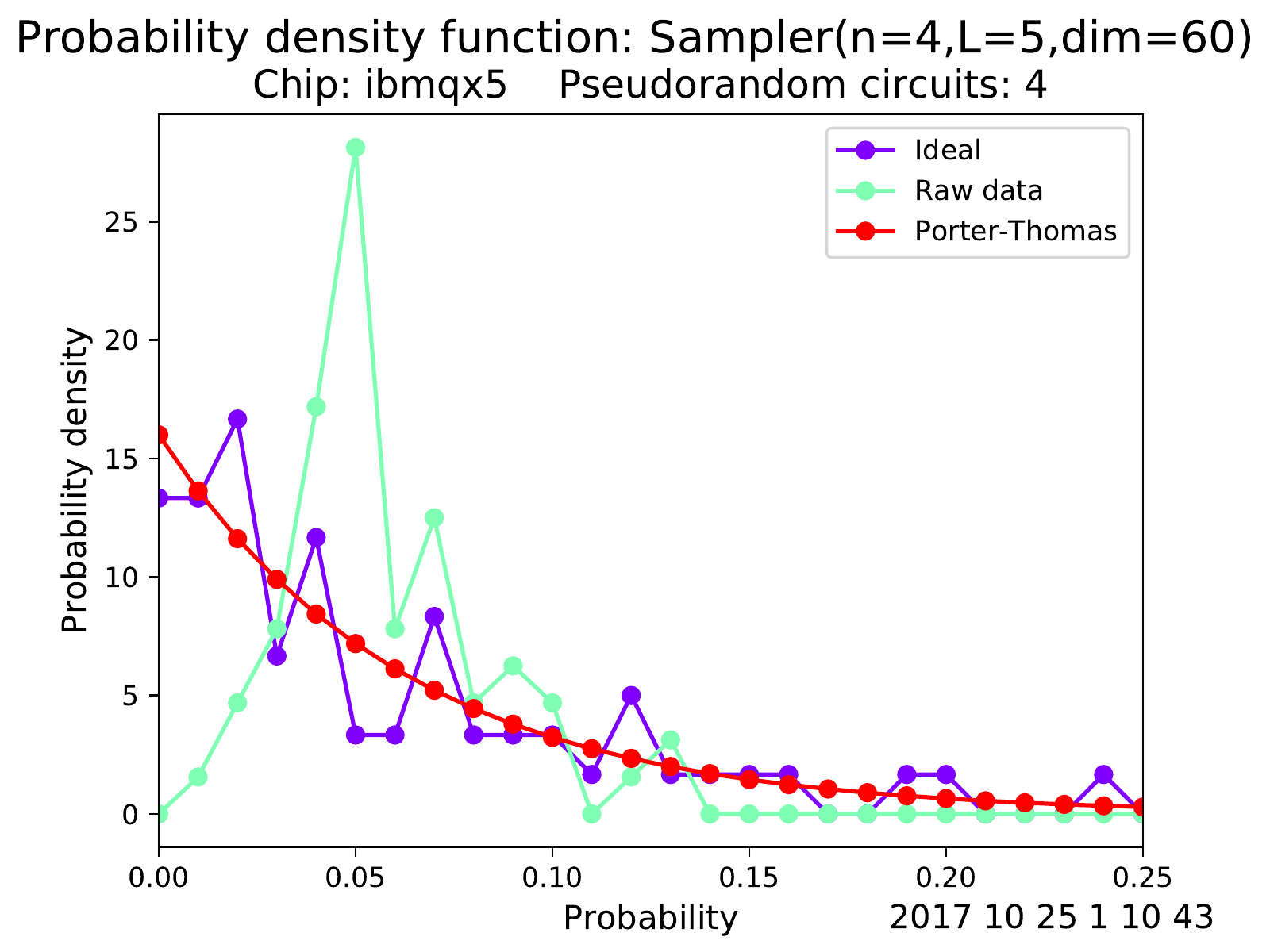} 
\caption{(Color online) Probability density function averaged over 4 pseudorandom samplers. The ideal data results from classical simulation of the circuits with no decoherence or gate errors. Here $1/N = 0.0625.$}
\label{pdf45}
\end{figure}

\begin{figure}
\includegraphics[width=8.5cm]{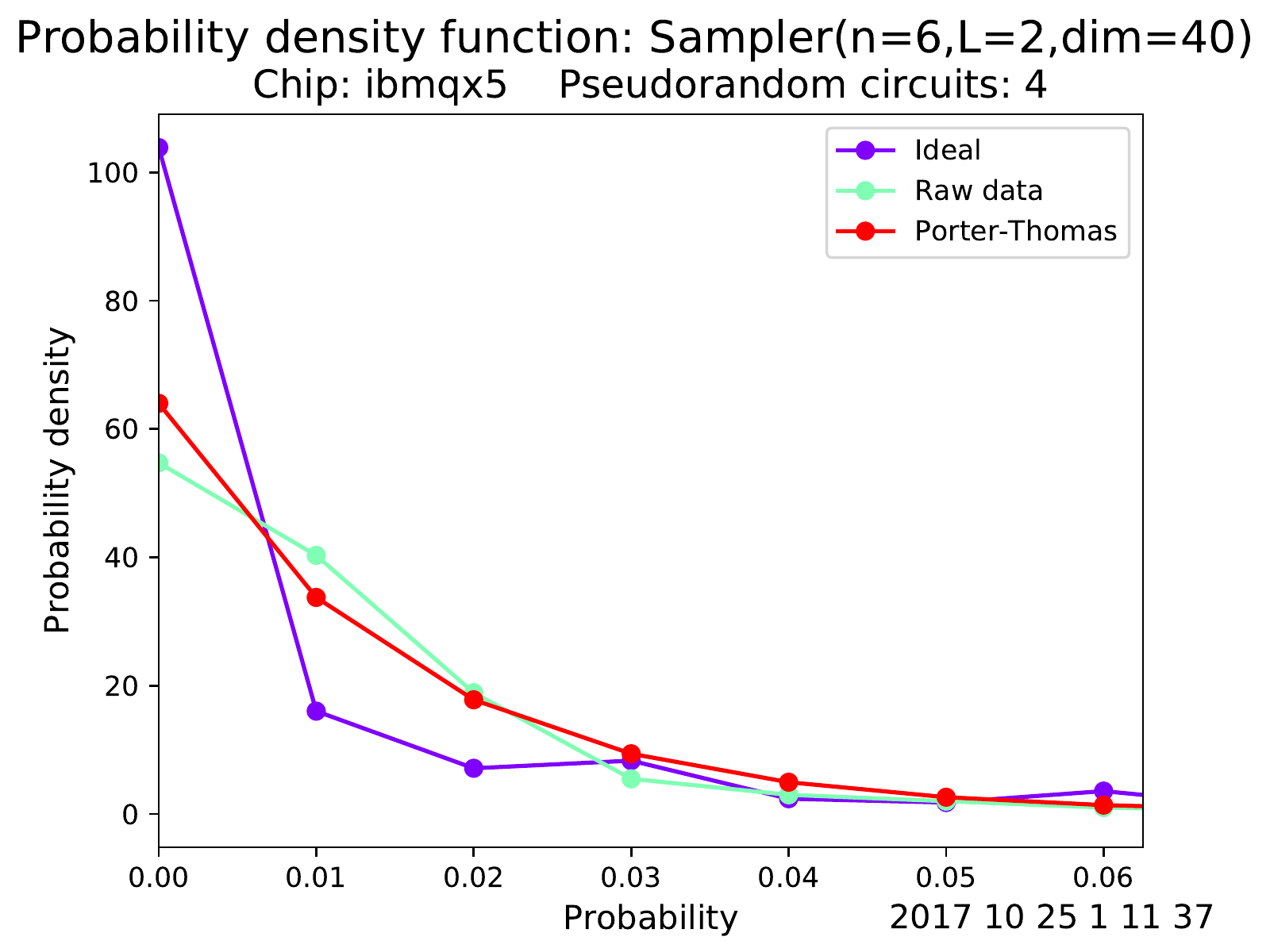} 
\caption{(Color online) Probability density function averaged over 4 pseudorandom samplers. In this case $1/N = 0.0156.$}
\label{pdf62}
\end{figure}

\begin{figure}
\includegraphics[width=8.5cm]{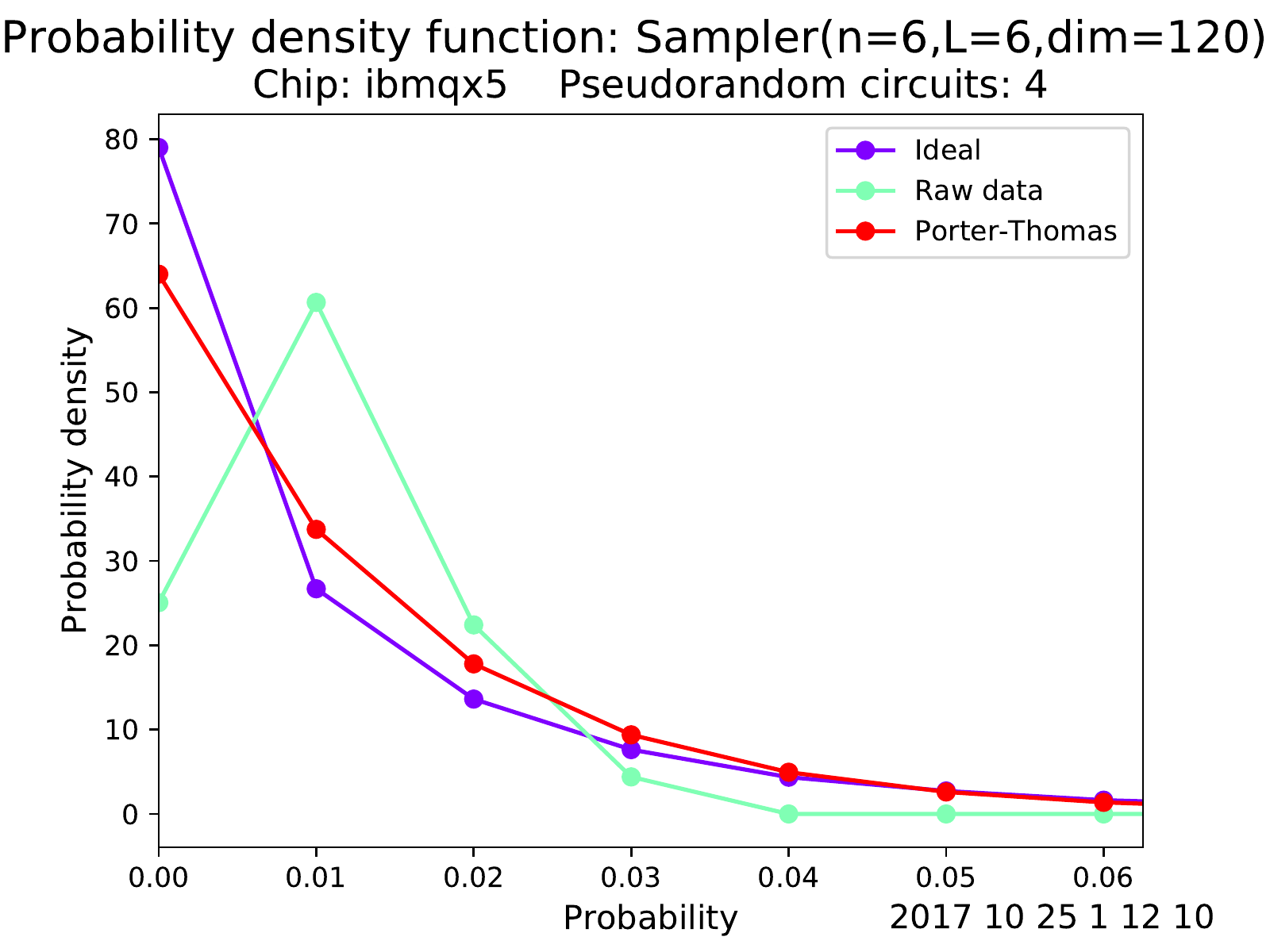} 
\caption{(Color online) Probability density function averaged over 4 pseudorandom samplers. $1/N = 0.0156.$}
\label{pdf66}
\end{figure}

A better approach is to use the fact that Haar typical unitaries are quantum chaotic \cite{Hayden0407049, Boixo160800263}, and to probe that chaos.
First we use the technique of Boixo {\it et al.}~\cite{Boixo160800263} and Neill {\it et al.}~\cite{Neill170906678}, and measure quantum chaos by its effect on the statistics of the sampled probability amplitudes. Probability density functions are shown in Figs.~\ref{pdf45} through \ref{pdf66}. For $n\!=\!4$ and smaller, the data are very noisy, but show the main features that we find for larger $n$: There is good agreement for probabilities larger than $1/N$, but that the frequency of small probabilities are suppressed relative to Porter-Thomas. Furthermore, we do not find a clear convergence to Porter-Thomas as a function of $L$. The agreement is already quite good after two layers, as shown in Fig.~\ref{pdf62} for six qubits, but it then declines as low-probability amplitudes become less frequent, as in Fig.~\ref{pdf66}. 

\begin{figure}
\includegraphics[width=8.5cm]{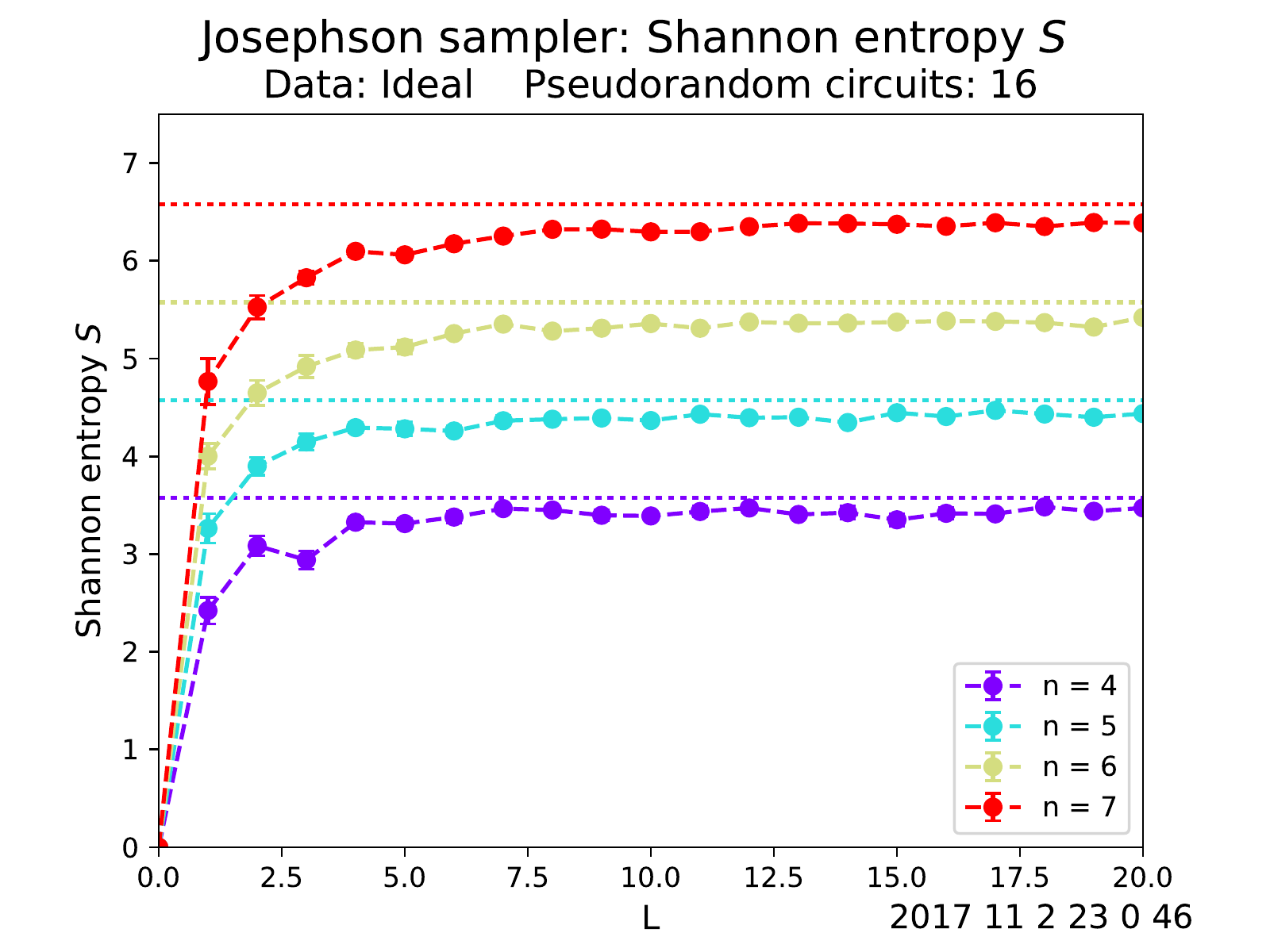} 
\caption{(Color online) Measured (solid curves) and simulated (dashed) entropy versus number of sampler layers. The dotted horizontal lines are the Porter-Thomas values (\ref{Spt}). In this plot we averaged over 16 pseudorandom samples to allow a better comparison with (\ref{Spt}). }
\label{entropyLarge}
\end{figure}

Porter-Thomas statistics is also reflected in the classical entropy taking the universal value \cite{Boixo160800263}
\begin{equation}
S_{\rm PT} = n -1 + \gamma 
\label{Spt}
\end{equation}
where $\gamma \approx 0.577$ is the Euler constant. In Fig.~\ref{entropyLarge} we show the ideal sampler entropy converging to (almost) the universal values. Note that PT-chaotic circuits are not maximally randomizing, as their classical entropy generation is short by 0.423 bits. However classical entropy measurement suffers from the same problem that $S_{\rm e}$ and $Q$ do.

Probing quantum chaos via Porter-Thomas statistics is probably best done with many circuit samples, and is not ideally suited for single-instance typicality testing. An alternative approach is to probe the butterfly effect, a dynamical manifestation of chaos. In the original setting \cite{Maldacena150301409} we consider a time-independent Hamiltonian $H$ and a pair of local, spatially non-overlapping (hence commuting) Hermitian observables $V$ and $W$. Consider the overlap between states
\begin{equation}
|a\rangle = W(t) V(0) |x \rangle
\ \ \ {\rm and} \ \ \ 
|b\rangle = V(0) W(t) | x \rangle,
\end{equation}
that only differ in the order that $V$ and $W$ are applied. 
Here $ W(t) = U^\dagger(t) W U(t),$ with $U(t) = e^{-iHt/\hbar}$ the time-evolution operator, and $| x \rangle$ is some initial state (in our case it is a classical state $x \in \{0,1\}^{\otimes n}$). The operation $W(t)$ propagates the state forward in time, applies $W$, and then propagates backwards in time. The states $|a\rangle$ and $|b\rangle$ differ by whether $V$ is applied before this excursion or after it. If $t \! = \! 0$, $|a\rangle$ and $|b\rangle$ are identical, because $V$ and $W$ commute. The key idea---or definition---is that chaotic dynamics in $U(t)$ will make them fail to commute at later times, resulting in a decrease of the overlap $|\langle a | b \rangle|^2,$ which we can measure through the 4-point correlation function
\begin{equation}
\big| \langle x | W(t) V(0) W(t) V(0) | x \rangle \big|^2 = |\langle a | b \rangle|^2.
\label{overlap}
\end{equation}

We will probe the chaos generated by a Josephson sampler circuit by using it place of the time evolution, using the inverse circuit for the reverse direction. For observables we use single-qubit Paulis
$Y_{1}$ and $X_{n}$ sitting at the ends of the chain, to mimic the classical limit $(V, W) \rightarrow (p, x).$ The butterfly effect we consider is defined by a nonvanishing of 
\begin{equation}
C = {\rm Tr} ([W, V]^\dagger  [W, V]) / N,
\end{equation}
where
\begin{equation}
V = Y_1,
\ \ \
W = U^\dagger X_{n} U,
\end{equation}
with $U$ and $U^\dagger$ provided by the sampler. Expanding the commutators leads to $C = 2(1-F),$ where
\begin{equation}
F = \frac{ {\rm Tr} \, W V W V}{N}  
= \frac{1}{N} \sum_{x=0}^{N-1} G(x) 
\label{Fdefinition}
\end{equation}
is real and
\begin{equation}
G(x) \equiv  \langle x | W V W V | x \rangle
\label{GxDefinition}
\end{equation}
is complex.
Here $G(x)$ gives the contribution from computational basis state $x \in \{0,1\}^{\otimes n}$ to the trace $F$.
Note that $|G(x)|^2$ is the 4-point function and overlap discussed in (\ref{overlap}). $F$ has the standard form of an out-of-time-order correlator (OTOC) used to diagnose the butterfly effect,  but with the time-evolution operator replaced by the Josephson sampler circuit. We will measure the absolute-value $|F|$, which the butterfly effect causes to vanish. An OTOC was also measured recently in a
nuclear magnetic resonance quantum simulator \cite{LiPRX17}.

\begin{figure}
\includegraphics[width=8.5cm]{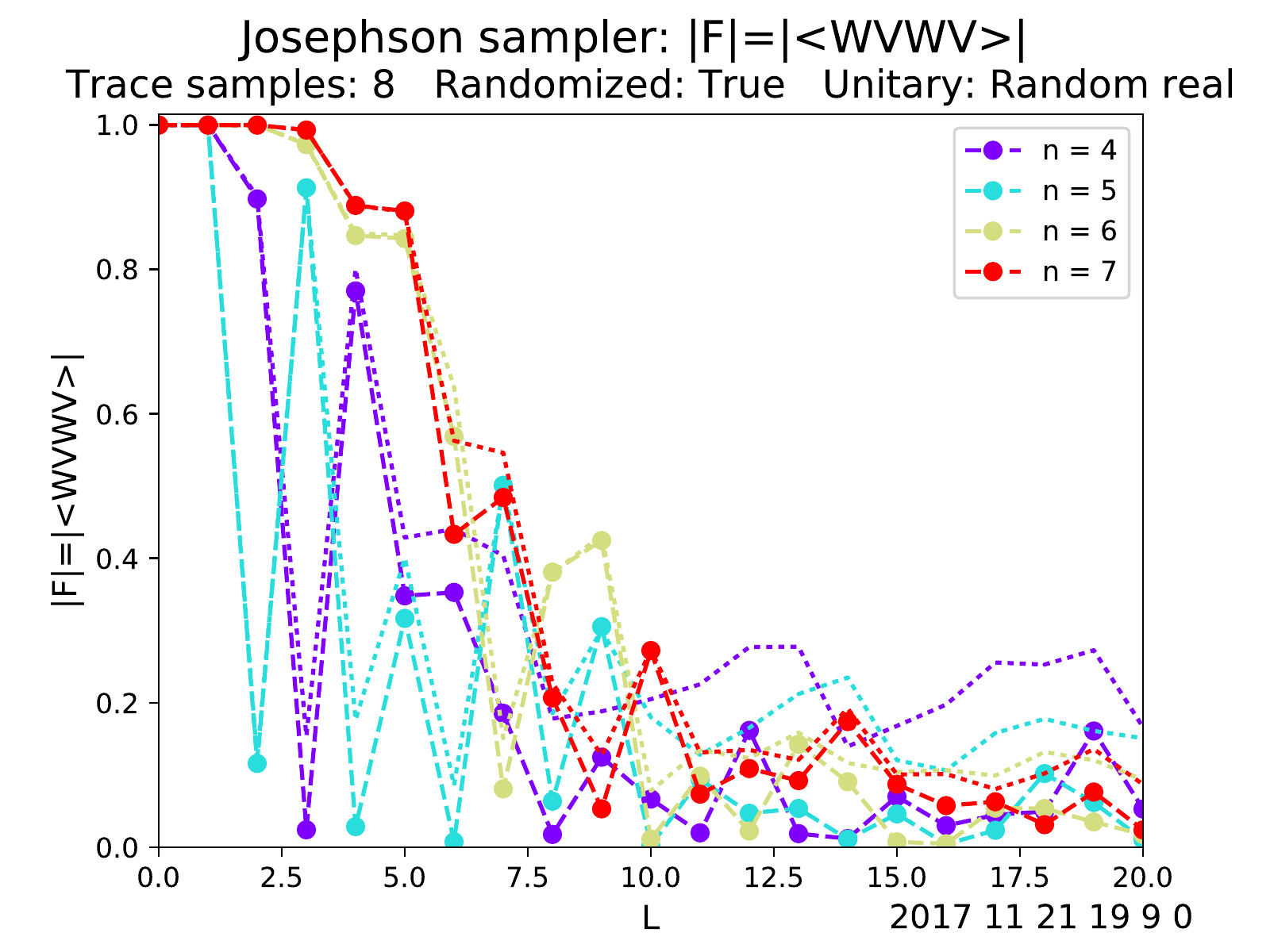} 
\caption{(Color online) Classically simulated OTOCs for the pseudorandom Josephson sampler. Dashed curves are exact, dotted curves use approximation (\ref{Fest}). }
\label{WVWVRandomRealLarge}
\end{figure} 

\begin{figure}
\includegraphics[width=8.5cm]{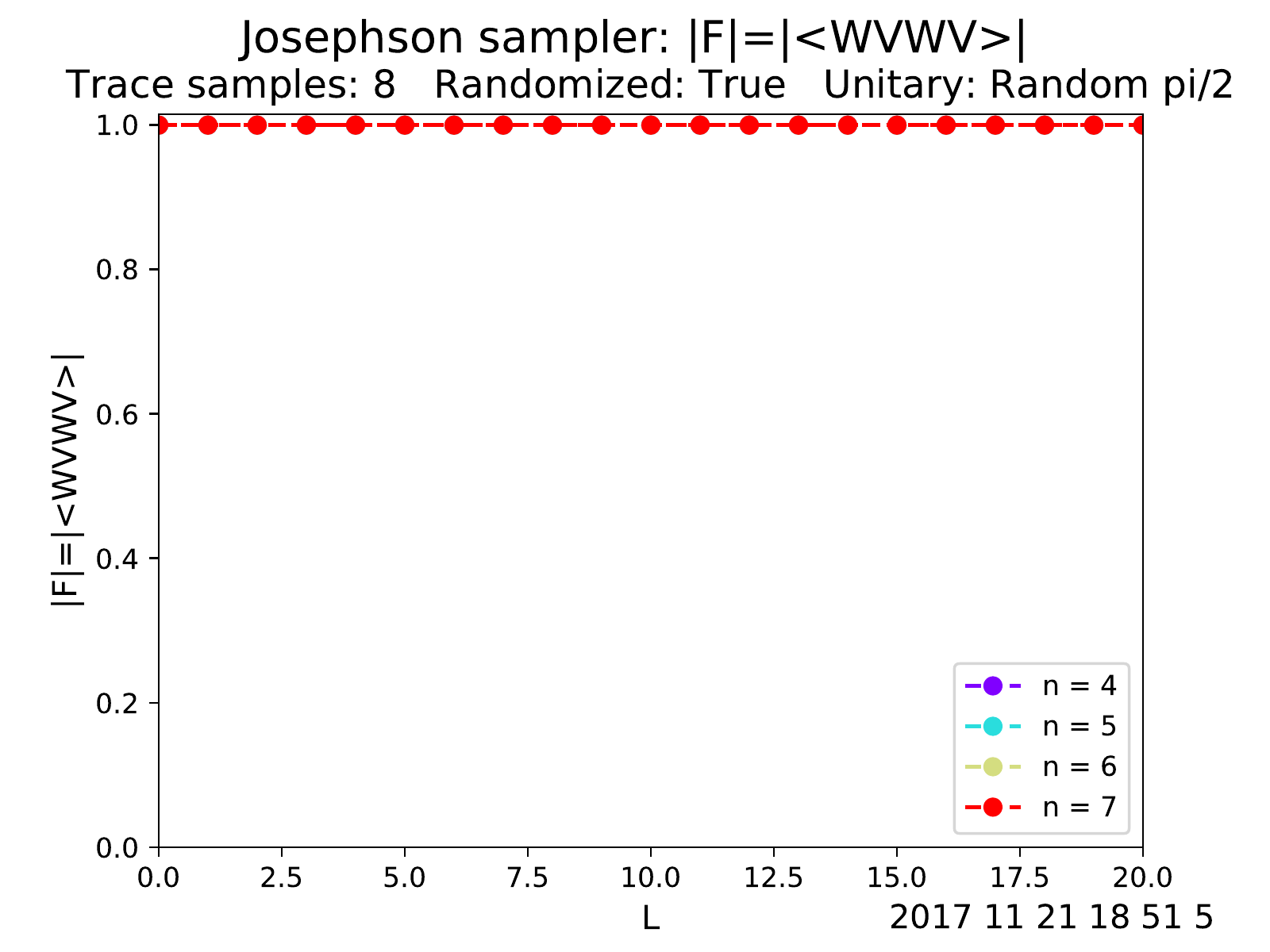} 
\caption{(Color online) Classically simulated OTOCs for the pseudorandom Josephson sampler with Clifford gates only. Dashed curves are exact, dotted curves use approximation (\ref{Fest}). Neither show scrambling (all curves are 1).}
\label{WVWVRandomPi2Large}
\end{figure} 

The measurement of $F$ would appear to require the measurement of $N \! = \! 2^n$ individual $G(x)$ correlators, one for each computational basis state $x$. But in the cases studied here, the real part of the complex quantity $G(x)$ is essentially independent of $x$, allowing us to estimate the trace $F$ with a small set $\{x' \} $ of $\nu$ classical states chosen at random from $\{0,1\}^{\otimes n}$,
\begin{equation}
F \approx \frac{1}{\nu} \sum_{x \in \{x' \} }  {\rm Re} \, G(x) .
\label{Festimation}
\end{equation}
Because $ {\rm Re} \, G(x)$ is approximately independent of $x$,
\begin{equation}
|F| \approx \frac{1}{\nu} \sum_{x \in \{x' \} } \big| {\rm Re} \, G(x) \big|.
\label{GabsEstimation}
\end{equation}
In addition to the stochastic trace evaluation, we introduce a second approximation, by ignoring the imaginary part of $G(x)$. While the real part of $G(x)$ is essentially $x$-independent and changes in magnitude from 1 to 0 as scrambling develops, the imaginary part is small (typically $< 0.10$ in the examples studied here) and random. Therefore we use
\begin{equation}
|F| \approx \frac{1}{\nu} \sum_{x \in \{x' \} } \big| G(x) \big|.
\label{Fest}
\end{equation}
Swingle {\it et al}. \cite{SwinglePRA16} discussed a similar approximation. 

\begin{figure}
\includegraphics[width=8.5cm]{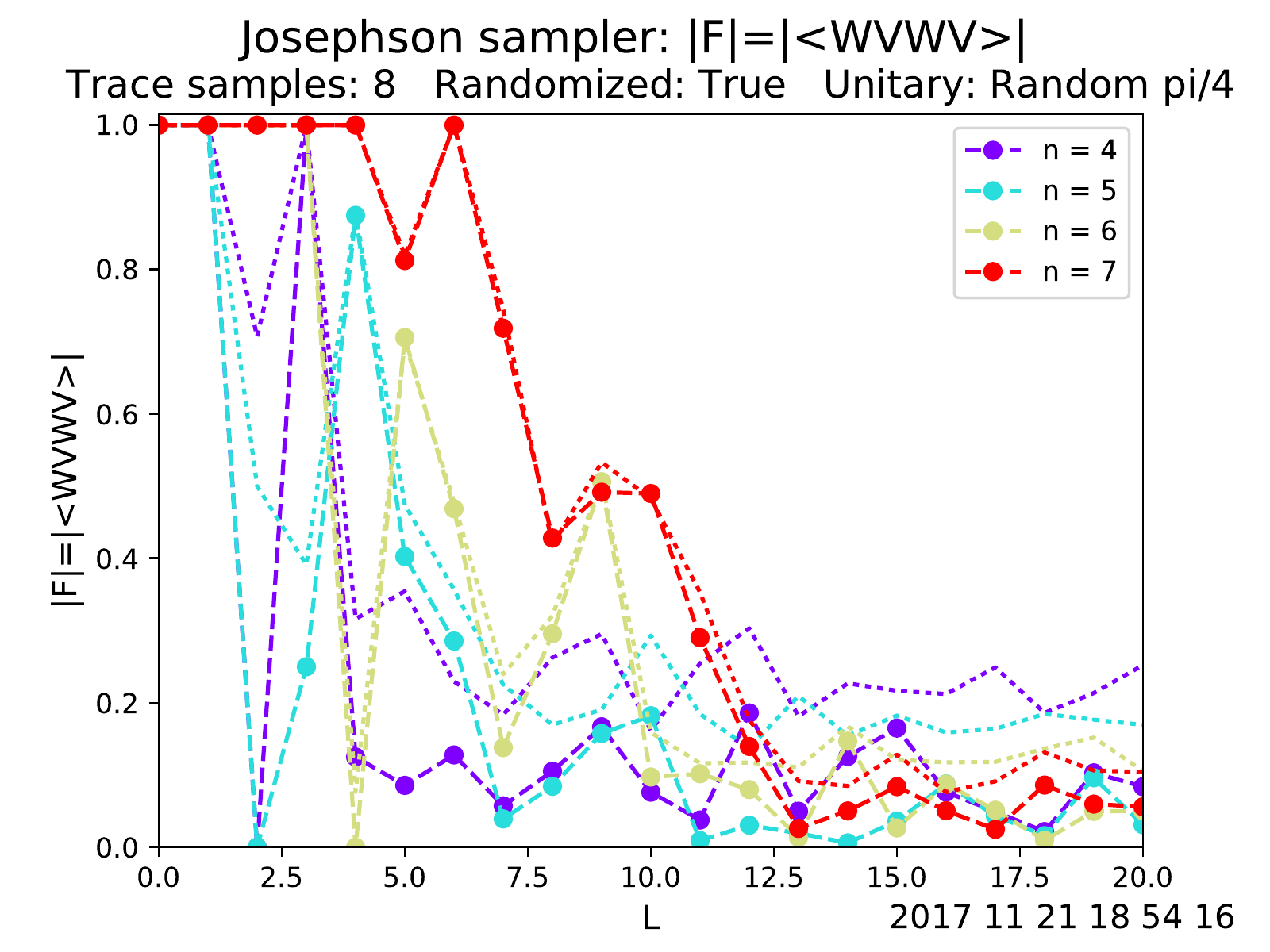} 
\caption{(Color online) This is the same as Fig.~\ref{WVWVRandomPi2Large}, but with single-qubit rotation angles given by random multiples of $\pi/4$.}
\label{WVWVRandomPi4Large}
\end{figure} 

Before discussing data, we will validate the approximation (\ref{Fest}). In Fig.~\ref{WVWVRandomRealLarge} we show classically simulated OTOCs for the pseudorandom Josephson sampler. The input data is a vector of pseudorandom real numbers in the range $[-2\pi,2\pi]$. The dashed curves show exact values of $F$ calculated from (\ref{Fdefinition}). The dotted curves are the ideal results within the approximation (\ref{Fest}), and we see that they easily diagnose the pronounced effect of scrambling in the chaotic regime. 
A second example, Fig.~\ref{WVWVRandomPi2Large}, uses a {\em Clifford sample}r, i.e., a pseudorandom Josephson sampler with only Clifford gates. To realize this we input a data vector consisting of random multiples of $\pi/2$ between $-2\pi$ and  $2\pi$. In this example the sampler does not exhibit OTOC decay. Although the scrambling and OTOC decay by unitary $k$-designs is a topic of current investigation, 4-designs are expected to be sufficient for OTOC decay \cite{Cotler170605400}, whereas the ideal Clifford sampler should be a 3-design \cite{Webb151002769}, suggesting that $k \! \ge \! 4$ is actually necessary. Including random non-Clifford gates, by instead using multiples of $\pi/4$, again leads to scrambling, shown in Fig.~\ref{WVWVRandomPi4Large}.

\begin{figure}
\includegraphics[width=8.5cm]{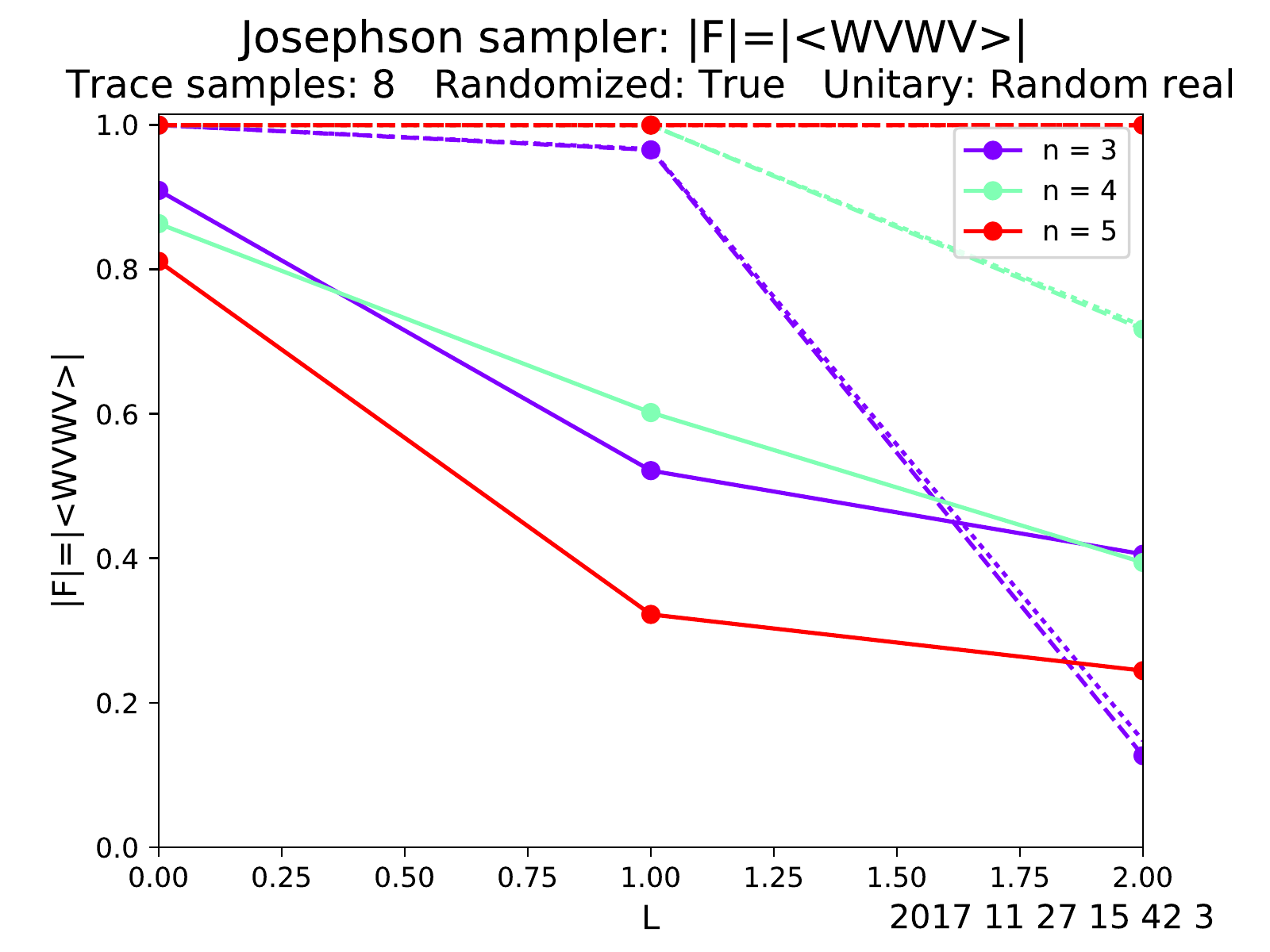} 
\caption{(Color online)  Measured (solid curves) and classically simulated (dashed) OTOCs for the pseudorandom Josephson sampler. Dotted curves show the ideal result within the approximation (\ref{Fest}).}
\label{WVWVRandomReal}
\end{figure} 

\begin{figure}
\includegraphics[width=8.5cm]{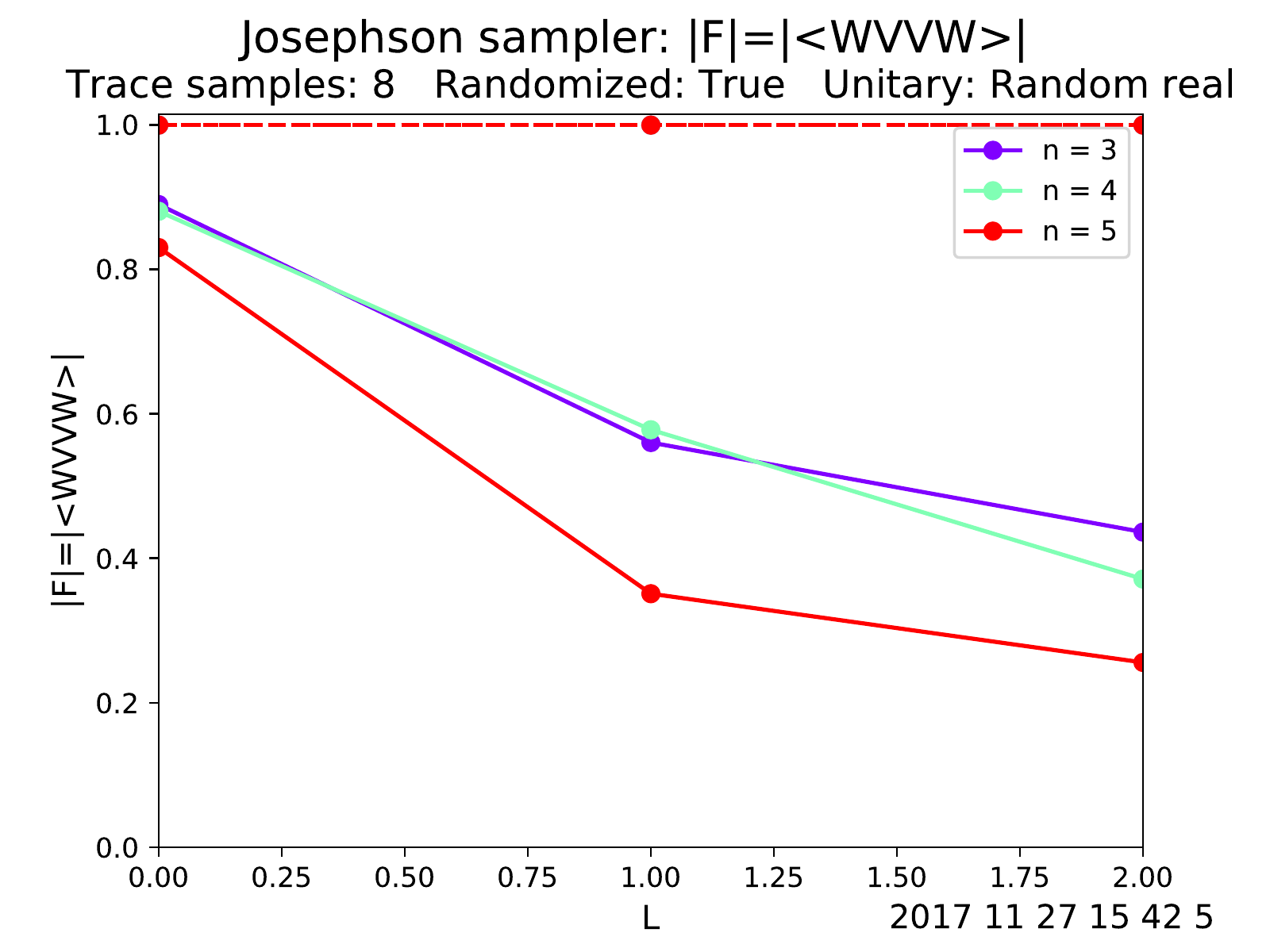} 
\caption{(Color online)  Measured (solid curves) and classically simulated (dashed) OTOCs  for the pseudorandom Josephson sampler. Dotted curves show the ideal result within the approximation (\ref{Fest}).}
\label{WVVWRandomReal}
\end{figure} 

These simulation results give evidence that a
rather small and shallow sampler circuit is capable of generating Haar random unitaries of sufficient quality as to exhibit quantum chaos and scrambling. However the measurement of each $|G(x)|$ requires a large, complex circuit that, except for very small cases, exceeds the current size limit of the Quantum Experience API.  We are able to measure OTOCs for samplers up to $n \! = \! 5$ and $L \! = \! 2$, as shown in Fig.~\ref{WVWVRandomReal}. The data are consistent with scrambling, but how do we distinguish genuine unitary scrambling from simple gate errors and decoherence? 

\begin{figure}
\includegraphics[width=8.5cm]{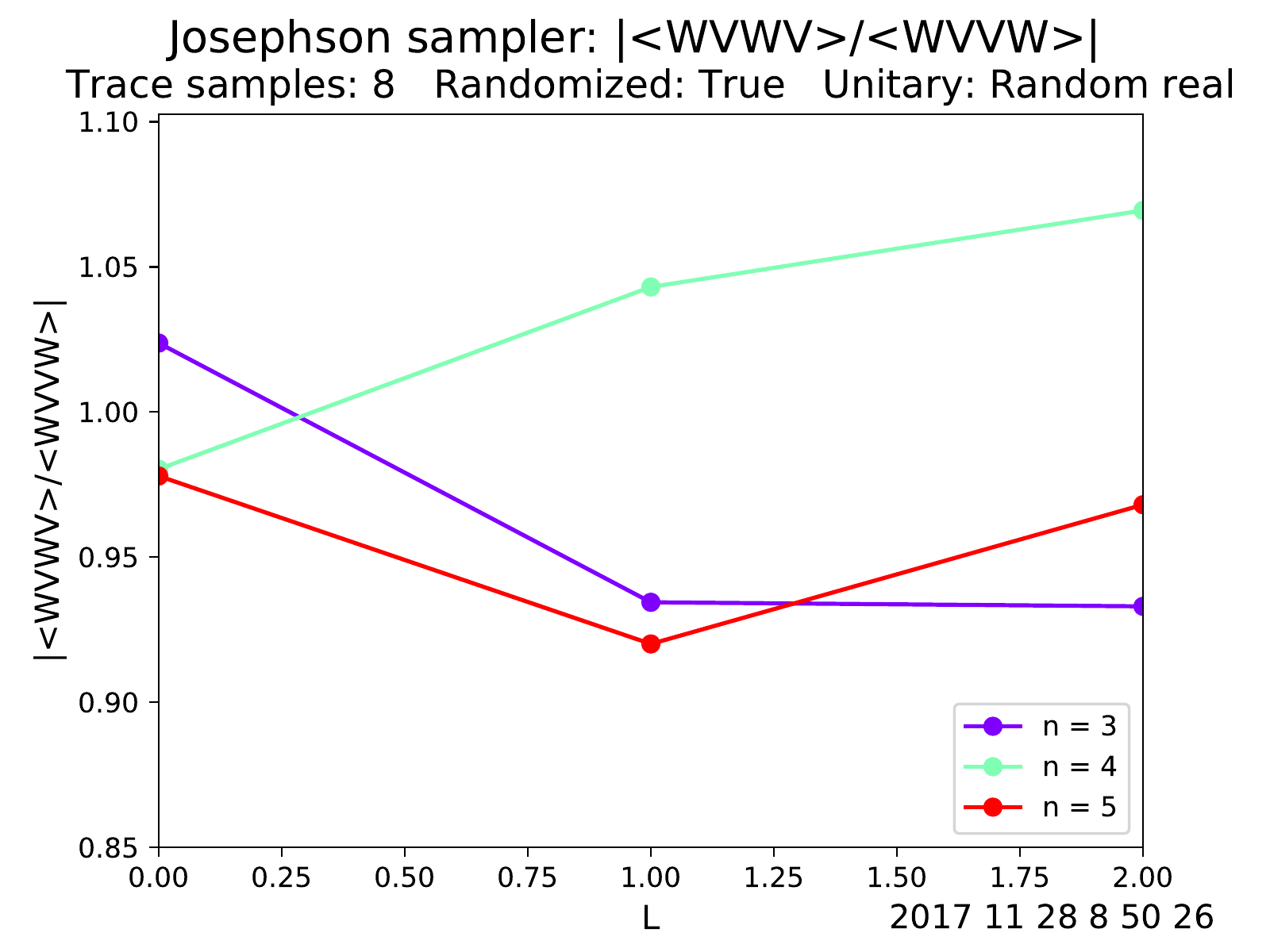} 
\caption{(Color online) Normalized (solid curves) OTOCs for the pseudorandom Josephson sampler.}
\label{RatioRandomReal}
\end{figure} 

In Fig.~\ref{WVVWRandomReal}
we plot the measured 4-point correlator $\langle WVVW \rangle$, which is ideally equal to one, and differs from (\ref{Fdefinition}) only in the order of the final 2 operators ($VW$ versus $WV$). In the presence of genuine unitary scrambling, we would expect the decay in $\langle WVWV \rangle$ to be faster than in $\langle WVVW \rangle$, and in Fig.~\ref{RatioRandomReal} we plot the ratio of these quantities, for a heuristic error-resistant metric. We conclude that the circuit fidelity is not yet sufficient to observe genuine unitary scrambling, but that it is not far off, giving support to the Haar typicality conjecture.

\begin{acknowledgments}

Data was taken on the IBM qx5 chip using the Quantum Experience API and the BQP software package developed by the author. The complete data set represented here consists of $\sim$3000 distinct circuits, each measured 8000 times. I'm grateful to IBM Research and the IBM Quantum Experience team for making their devices available to the quantum computing community. I also want to thank Sergio Boixo, Jerry Chow, Jordan Cotler, Andrew Cross, Charles Neill, Hanhee Paik, and Beni Yoshida for their private communication. Thanks also to Amara Katabarwa, Mingyu Sun, Jason Terry, and Phillip Stancil for their discussions and contributions to BQP. This work does not reflect the views or opinions of IBM or any of its employees. 
\end{acknowledgments}

\newpage

\appendix

\section{JOSEPHSON SAMPLER CIRCUIT}

\begin{figure}
\includegraphics[width=7.5cm]{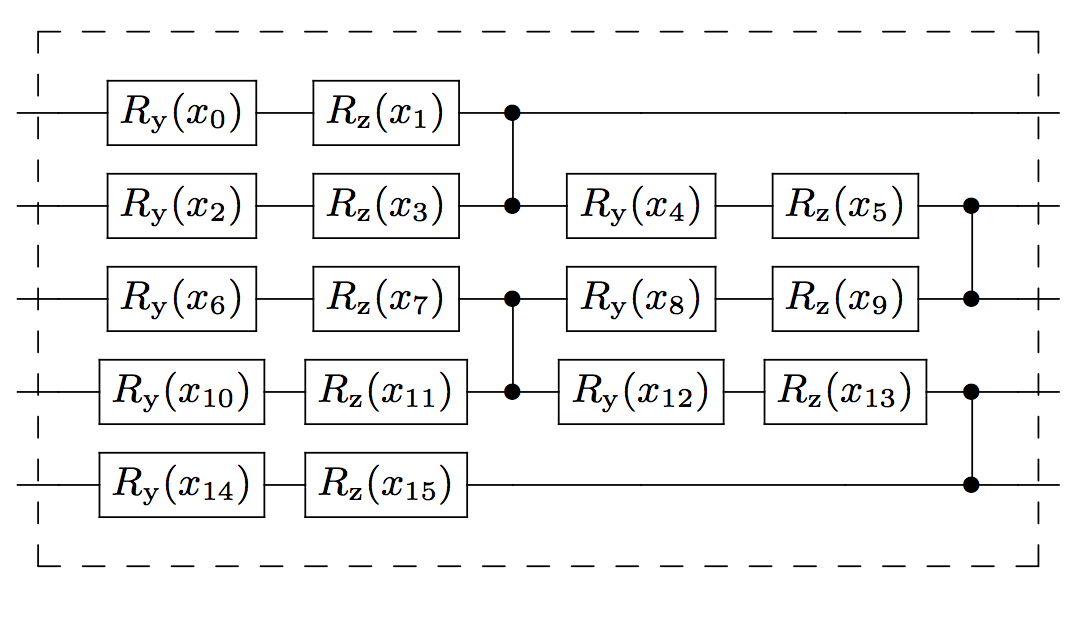} 
\caption{(Color online) $n\!=\!5$ Josephson sampler layer.}
\label{n5}
\end{figure} 

\begin{figure}
\includegraphics[width=7.5cm]{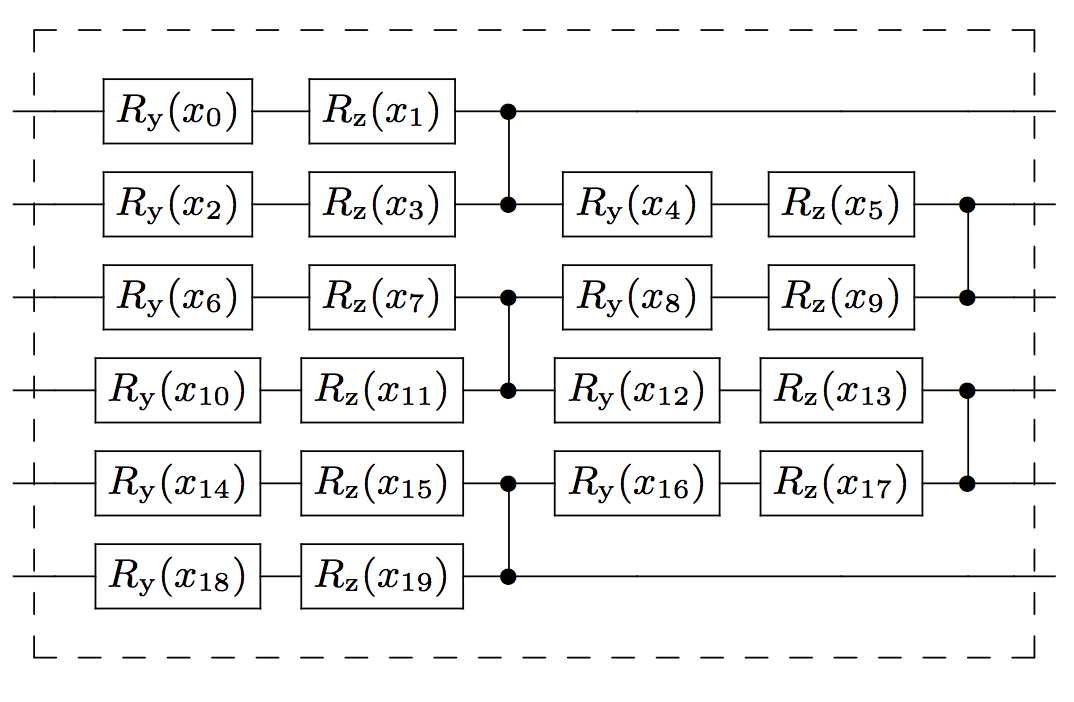} 
\caption{(Color online) $n\!=\!6$ Josephson sampler layer.}
\label{n6}
\end{figure}

Here we provide additional details about the Josephson sampler. In (\ref{u definition}), $Y$ and $Z$ are Pauli matrices. The circuit is a function of $x \in \mathbb{R}^m$, with each component $4\pi$-periodic. The design attempts to embed as many rotation parameters as possible, while making sure there is no redundancy when applied to classical inputs or when layered. On the imbqx5 chip the CZ gate ${\rm diag}(1,1,1,-1)$ is made from a CNOT and Hadamards. Explicit $L \! = \! 1$ circuits for $n\!=\!5$ and 6 are shown in Figs.~\ref{n5} and \ref{n6}, which also show the particular mapping between vector components and gate angles used.

\bibliography{/Users/mgeller/Dropbox/bibliographies/algorithms,/Users/mgeller/Dropbox/bibliographies/applications,/Users/mgeller/Dropbox/bibliographies/dwave,/Users/mgeller/Dropbox/bibliographies/control,/Users/mgeller/Dropbox/bibliographies/error_correction,/Users/mgeller/Dropbox/bibliographies/general,/Users/mgeller/Dropbox/bibliographies/group,/Users/mgeller/Dropbox/bibliographies/ions,/Users/mgeller/Dropbox/bibliographies/math,/Users/mgeller/Dropbox/bibliographies/ml,/Users/mgeller/Dropbox/bibliographies/nmr,/Users/mgeller/Dropbox/bibliographies/optics,/Users/mgeller/Dropbox/bibliographies/simulation,/Users/mgeller/Dropbox/bibliographies/superconductors,/Users/mgeller/Dropbox/bibliographies/surfacecode}

\end{document}